\documentclass[notitlepage,aps,longbibliography,letter,reprint,nofootinbib,superscriptaddress]{revtex4-1}

\usepackage{graphicx}
\usepackage{amsmath}
\usepackage{amssymb}
\usepackage{hyperref}
\usepackage{color}
\usepackage[abs]{overpic}
\usepackage[english]{babel}

\usepackage{letltxmacro}

\LetLtxMacro{\ORIGselectlanguage}{\selectlanguage}
\makeatletter
\DeclareRobustCommand{\selectlanguage}[1]{%
  \@ifundefined{alias@\string#1}
    {\ORIGselectlanguage{#1}}
    {\begingroup\edef\x{\endgroup
       \noexpand\ORIGselectlanguage{\@nameuse{alias@#1}}}\x}%
}
\newcommand{\definelanguagealias}[2]{%
  \@namedef{alias@#1}{#2}%
}
\makeatother

\definelanguagealias{en}{english}

\newcommand{\be}{\begin{eqnarray*}}
\newcommand{\ee}{\end{eqnarray*}}
\newcommand{\beq}{\begin{eqnarray}}
\newcommand{\eeq}{\end{eqnarray}}
\newcommand{\bequ}{\begin{equation}}
\newcommand{\eequ}{\end{equation}}
\newcommand{\bk}{{\mathbf{k}}}

\newcommand{\bB}{{\mathbf{B}}}

\newcommand{\dd}{\mathrm{d}}

\newcommand{\h}{\hat{H}}
\newcommand{\up}{\uparrow}
\newcommand{\down}{\downarrow}
\newcommand{\ph}{{\phantom{\dagger}}}
\newcommand{\ket}[1]{\left|{#1}\right\rangle}
\newcommand{\bra}[1]{\left\langle{#1}\right|}
\newcommand{\hc}{\mathrm{H.c.}}

\begin{document}
\title{Majorana Braiding in Realistic Nanowire Y-Junctions and Tuning Forks}
\author{Fenner~Harper}
\affiliation{Mani L. Bhaumik Insitute for Theoretical Physics, Department of Physics and Astronomy, UCLA, Los Angeles CA 90095, USA}
\author{Aakash~Pushp}
\affiliation{IBM Almaden Research Center, San Jose, California 95120, USA}
\author{Rahul~Roy}
\affiliation{Mani L. Bhaumik Insitute for Theoretical Physics, Department of Physics and Astronomy, UCLA, Los Angeles CA 90095, USA}
\date{\today}
\begin{abstract}
Majorana fermions are predicted to arise at the ends of nanowire devices which combine superconductivity, strong spin-orbit coupling and an external magnetic field. By manipulating networks of these devices with suitable gating, it has been suggested that braiding operations may be performed which act as logic operations, suitable for quantum computation. However, the unavoidable misalignment of the magnetic field in any realistic device geometry has raised questions about the feasibility of such braiding. In this paper, we numerically simulate braiding operations in devices with Y-junction and tuning fork geometries using an experimentally motivated nanowire model. We study how the static and dynamical features vary with geometric parameters and identify parameter choices that optimise the probability of a successful braid. Notably, we find that there is an optimal Y-junction half-angle (about $20^\circ$ for our parameter values), which balances two competing mechanisms that reduce the energy gap to excitations. In addition, we find that a tuning fork geometry has significant advantages over a Y-junction geometry, as it substantially reduces the effect of dynamical phase oscillations that complicate the braiding process. Our results suggest that performing a successful braid is in principle possible with such devices, and lies within experimental reach.
\end{abstract}
\maketitle

\section{Introduction}
Bound states of Majorana fermions are believed to exhibit non-abelian statistics \cite{AliceaNew2012,StanescuMajorana2013,BeenakkerSearch2013,SarmaMajorana2015,ElliottColloquium2015,AasenMilestones2016,AguadoMajorana2017,LutchynMajorana2018,ZhangQuantum2019}, and so offer the exciting possibility of realising a topological quantum computer \cite{NayakNonAbelian2008,KitaevFaulttolerant2003}. This ambition, coupled with experimental advances, has focussed a great deal of recent effort towards realising emergent Majorana quasiparticles in condensed matter systems. Notably, suitable Majorana bound states are predicted to arise in certain fractional quantum Hall states \cite{MooreNonabelions1991}, in $p$-wave superconductors in one and two dimensions \cite{ReadPaired2000,KitaevUnpaired2001}, and in many other low-temperature solid state systems \cite{SauNonAbelian2010}. Beyond this, many varied theoretical proposals have been suggested which would allow the non-abelian properties of Majorana modes to be leveraged into a topological qubit \cite{AasenMilestones2016}. While compelling experimental evidence for the existence of Majorana modes has been detected in many of these settings (and in particular in nanowires \cite{MourikSignatures2012,DasZerobias2012,DengMajorana2016,ChenExperimental2017,NicheleScaling2017a,DengNonlocality2018,GulBallistic2018c,ShenParity2018a,OFarrellHybridization2018a,VaitiekenasFluxinduced2018,GrivninConcomitant2018,ZhangQuantized2018,vanZantenPhoton2019}), a completely unambiguous signature of Majorana modes remains lacking. In this work, we study the braiding of a pair of Majorana fermions in a realistic nanowire device, which, if reproduced experimentally, could provide one such unambiguous signature.

The area of Majorana nanowires is now fairly mature \cite{AguadoMajorana2017,LutchynMajorana2018,ZhangQuantum2019}, but much of our understanding of such systems stems from Kitaev's toy model of a $p$-wave superconducting chain \cite{KitaevUnpaired2001}. In the topological regime of this model, the ends of the wire host a pair of localised Majorana zero modes (MZMs), which may be moved around by adjusting the local system parameters. By forming networks of such wires, non-abelian braiding operations may be performed. The Kitaev model, however, requires exotic superconducting pairing, and is therefore difficult to realise in a real material. Instead, most experiments are believed to approximate the more realistic nanowire models of Refs.~\cite{SauNonAbelian2010,OregHelical2010,LutchynMajorana2010}. In these systems, strong spin-orbit coupling, $s$-wave superconducting pairing, and a moderately strong magnetic field conspire to produce the conditions necessary for MZMs. 

Nanowires based on this model have now been developed by several groups, both using direct epitaxial growth and by depleting regions of a two-dimensional electron gas to leave an effectively one-dimensional (1D) channel \cite{MourikSignatures2012,DasZerobias2012,DengMajorana2016,ChenExperimental2017,NicheleScaling2017a,DengNonlocality2018,GulBallistic2018c,ShenParity2018a,OFarrellHybridization2018a,VaitiekenasFluxinduced2018,GrivninConcomitant2018,ZhangQuantized2018,vanZantenPhoton2019}. Many of these experiments have reported transport signatures consistent with the existence of MZMs: notably, a MZM should lead to a robust zero-bias conductance peak quantised to $2e^2/h$ \cite{FlensbergTunneling2010,LawMajorana2009} (although disorder and finite-temperature effects may disguise this \cite{WimmerQuantum2011}). However, such a signature is not `smoking gun' evidence of a MZM: similar peaks can also be caused by spurious Andreev bound states, non-topological states close to zero energy which form due to local variations in the device parameters \cite{LiuAndreev2017,HellDistinguishing2018,MooreTwoterminal2018,LiuDistinguishing2018a,ReegZeroenergy2018a,AseevDegeneracy2019}. It has been suggested that more convincing evidence for Majorana modes could be obtained through interference experiments \cite{AkhmerovElectrically2009a,FuProbing2009,FuElectron2010,SauProposal2015}, by performing simultaneous tunnelling into the device at each end of the wire \cite{LiuAndreev2017}, or by studying spin-dependent transport signatures \cite{RiccoSpindependent2018}. Ultimately, however, the most direct way to probe the non-abelian nature of MZMs would be to perform a logic operation such as a braid. Such operations could form the basis of a topological quantum computer (albeit one that is not, on its own, universal \cite{BravyiUniversal2006}).

Experimentally, we might envisage performing a braid following the proposal of Ref.~\cite{AliceaNonAbelian2011}, where a series of side gates are attached to a three-legged device (or `Y-junction'), as in Fig.~\ref{fig:braiding}. By adjusting the voltage of the side gates, the local chemical potential may be changed, and different regions of the wire can be made topological or trivial as desired. If there are initially two pairs of MZMs, one Majorana from each pair can be braided following the protocol indicated in Fig.~\ref{fig:braiding}. Although the non-abelian nature of this operation has been confirmed in the ideal case \cite{AliceaNonAbelian2011}, there has been some doubt about its feasibility in a real material system. One concern is that the magnetic field (directed along the mother branch of a `Y-junction') is not parallel to the wire direction in the prongs of the device, which is known to reduce the size of the bulk gap \cite{MourikSignatures2012,OscaEffects2014,RexTilting2014,NijholtOrbital2016}. In addition, the effective superconducting pairing in the non-parallel legs can lead to the formation of a $\pi$-junction \cite{AliceaNonAbelian2011}. Beyond this, decoherence and quasiparticle poisoning times may be too short to allow the successful completion of a braid \cite{RainisMajorana2012a,HigginbothamParity2015,AlbrechtTransport2017,ZhangEffects2019,ClarkeProbability2017}.

In this paper, we numerically study the feasibility of such a braid in a realistic nanowire system. In previous works, braiding of Majoranas has been performed successfully in numerical simulations of the Kitaev model \cite{AmorimMajorana2015,SekaniaBraiding2017}, while braiding of non-Majorana defects has also been carried out in  simulations of the Su-Schrieffer-Heeger model \cite{BorossPoor2019}. In this work, we instead focus on the more realistic continuum nanowire model \cite{SauNonAbelian2010,OregHelical2010,LutchynMajorana2010}, which allows us to take into account the misalignment of the external magnetic field and the device geometry directly. As our starting point, we use parameters from existing state-of-the-art simulations of nanowire devices, which have been shown to reproduce experimental conductance data extremely well \cite{LiuAndreev2017,ZhangQuantized2018}. However, motivated by an ongoing experimental collaboration, we also incorporate material parameters and system geometries that are particularly relevant to devices grown using IBM's template-assisted selective epitaxy (TASE) technique \cite{BorgVertical2014,CzornomazConfined2015,GoothBallistic2017,SchmidTemplateassisted2015,GoothBallistic2017a}. While we do not claim to draw categorical conclusions about braiding in a specific (existing or proposed) device, we believe our results provide qualitative and approximate quantitative statements about braiding in devices which are within current experimental reach, and which overall are encouraging.

The structure of this paper is as follows. In Sec.~\ref{sec:majorana_nanowires} we introduce the nanowire model we will be using throughout this paper and discuss the ideal braiding properties of Majorana fermions. In Sec.~\ref{sec:static_properties} we study the static (instantaneous) properties of a nanowire Y-junction as the braid is performed, and identify geometric parameters which optimise the important energy scales of the system. In Sec.~\ref{sec:braiding_simulations} we study the dynamical properties of the system by performing numerical braiding simulations. In particular, we consider how the braiding success varies as parameters and time scales are modified. In Sec.~\ref{sec:tuning_forks} we discuss how using a tuning fork geometry offers several advantages over a Y-junction geometry, and investigate the robustness of our conclusions to variations in the underlying parameters. Finally, in Sec.~\ref{sec:conclusion} we summarise our results and provide some concluding remarks.

\section{Majorana Fermions in Nanowires\label{sec:majorana_nanowires}}
\subsection{Nanowire Models}
\begin{figure*}
\includegraphics[scale=0.3]{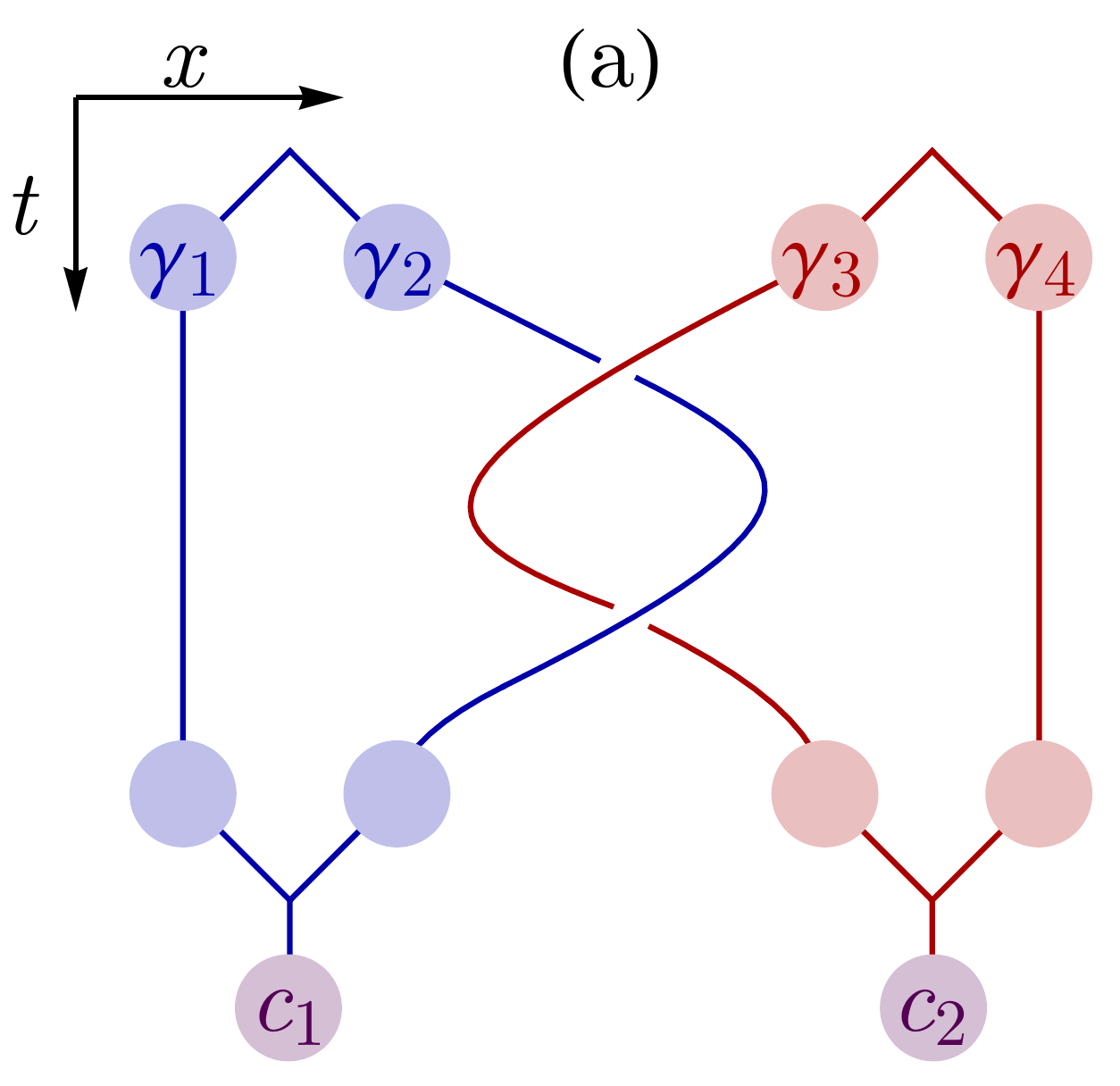}\hspace{7mm}
\includegraphics[scale=0.3]{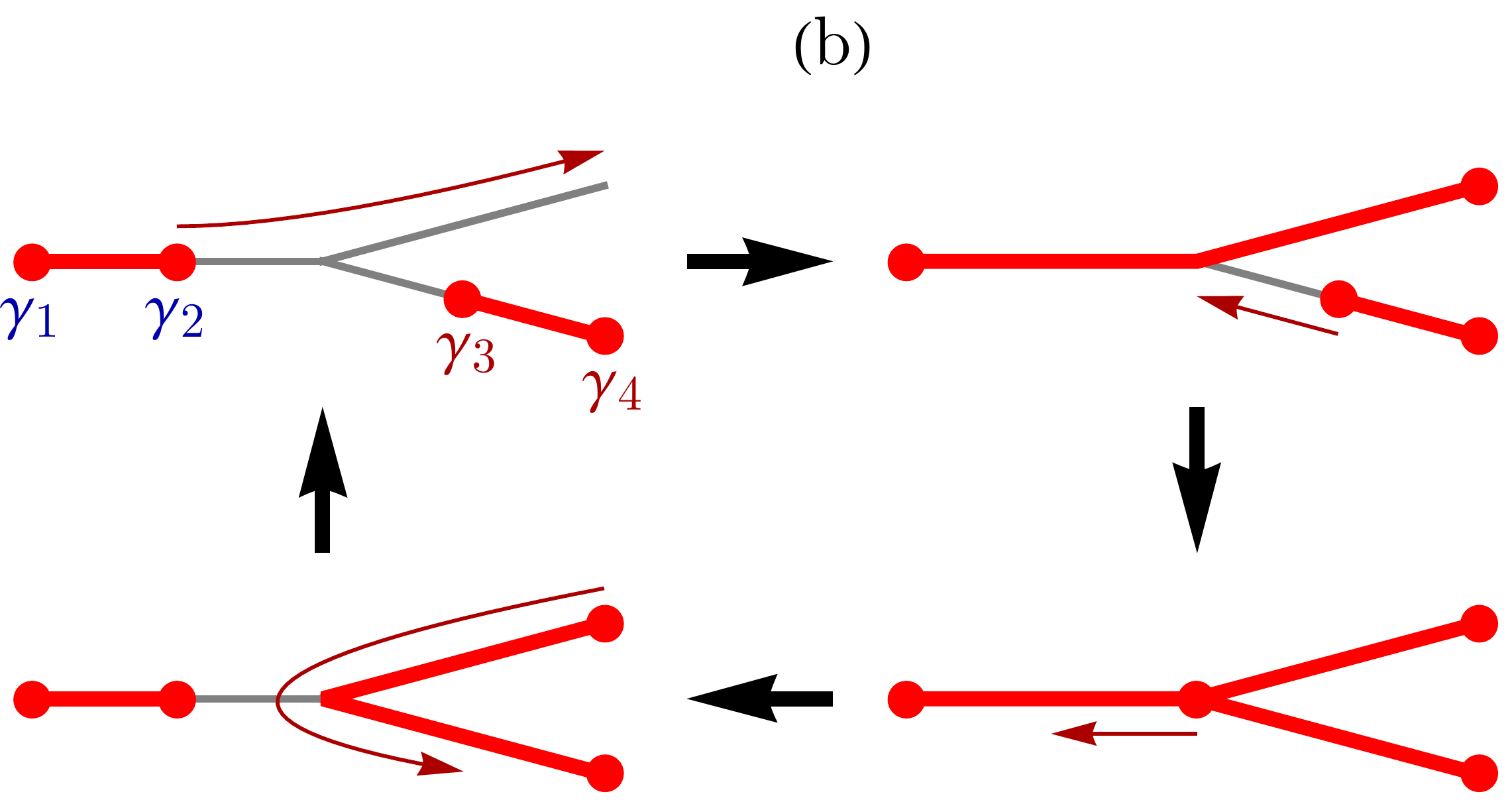}\hspace{5mm}
\includegraphics[scale=0.3]{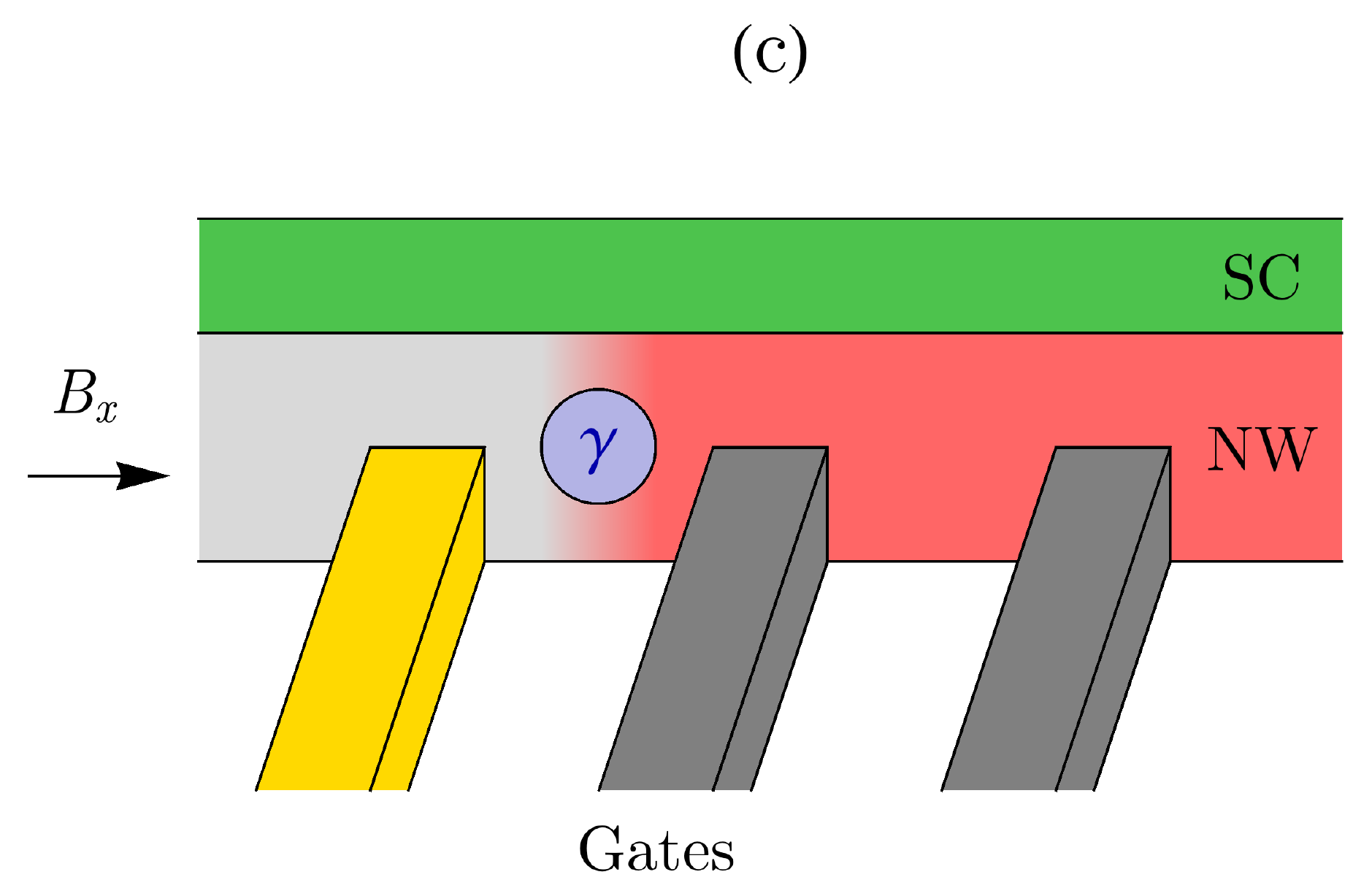}
\caption{(a) Ideal Majorana braid: two pairs of Majorana fermions ($\gamma_i$) are created from the vacuum, and one from each pair is braided. After this process, each pair of Majorana fermions fuses to form a complex fermion ($c_i$). (b) Sequence of moves for performing a braid in a nanowire Y-junction, starting in the top left. Red (thick) lines indicate sections of the wire in the topological regime, gray (narrow) lines indicate sections of the wire in the trivial regime, and red disks indicate Majorana fermions realised at the phase boundaries. The labels $\gamma_j$ indicate how the starting positions of the Majorana modes may be identified with those in panel~a. Note that this exchange must be carried out twice to perform a complete braid. (c) Schematic experimental setup for manipulating Majorana fermions in a real device. A section of a nanowire is shown (NW), with a superconducting layer grown on top (SC). When a moderate magnetic field is applied along the axis of the nanowire, the system may enter a topological superconducting phase. Side gates may be used (yellow/left switched on, gray/right switched off) to adjust the local electrostatic potential and thus change the location of the trivial regions (light gray/left) and topological regions (red/right). \label{fig:braiding}}
\end{figure*}

Majorana fermions arise in condensed matter systems as emergent fermionic quasiparticles which are their own `antiparticle'. Explicitly, Majorana fermions associated with operators $\gamma_i$ satisfy the relations $\gamma_i^\dagger=\gamma_i^\ph$ and $\left\{\gamma_i,\gamma_j\right\}=2\delta_{ij}$. They can always be defined (at least mathematically) by combining complex fermion creation and annihilation operators ($c^\dagger_j$ and $c^\ph_j$) through the relations
\beq
\gamma_{2j-1}=c^\dagger_j+c^\ph_j,&~~~~~~&\gamma_{2j}=i\left(c^\dagger_j-c^\ph_j\right),
\eeq
and in this way may be thought of as fractional excitations of the underlying electrons of a system. However, the transformation given above is not usually very useful---the resulting Majorana operators are often entangled in space and do not correspond to physically meaningful excitations of the system. Nevertheless, in the presence of superconductivity (where the particle number is only conserved modulo two and the quasiparticles are coherent superpositions of particles and holes) Majorana modes can arise as natural excitations. In particular, Majorana modes that are pinned to zero energy and which are spatially well-separated can arise as topological degrees of freedom.

The simplest 1D model which exhibits MZMs is the previously mentioned Kitaev chain \cite{KitaevUnpaired2001}, which has the Hamiltonian
\bequ
\h_{\rm K}=-\mu\sum_j\left[ c^\dagger_{j}c^\ph_j\right]-\frac{1}{2}\sum_j\left[tc^\dagger_{j}c^\ph_{j+1}+\Delta e^{i\phi}c^\ph_{j}c^\ph_{j+1}+\hc\right].\label{eq:Kitaev}
\eequ
In this Hamiltonian, $c^\dagger_j$ creates a spinless fermion on site $j$, $\mu$ is the chemical potential, $t>0$ describes the intersite hopping, and $\Delta e^{i\phi}$ describes the $p$-wave superconducting pairing. When written in terms of Majorana fermions, this Hamiltonian describes pairing between Majorana fermions on the same site and between Majorana fermions from neighbouring sites. If $\left|\mu\right|<t$, the system is in the topological regime, and an unpaired MZM arises at each end of the chain.
\footnote{We note that Eq.~\eqref{eq:Kitaev} is also related to the XY spin chain through the Jordan-Wigner transformation---see, for example, Ref.~\onlinecite{DerzhkoJordanWigner2008a}.}

Unfortunately, $p$-wave superconductivity is difficult to realise in the laboratory, but more realistic models which support MZMs were introduced in Refs.~\cite{SauNonAbelian2010,OregHelical2010,LutchynMajorana2010}. In these cases, the underlying system is a 1D semiconducting nanowire with strong Rashba spin-orbit coupling, usually corresponding experimentally to InAs or InSb. The nanowire is then proximity coupled to an $s$-wave superconductor, achieved experimentally by depositing a material such as Al as a thin layer on top of the nanowire, as shown schematically in Fig.~\ref{fig:braiding}(c). In zero magnetic field, a system of this kind is a trivial quasi-1D superconductor. However, if a magnetic field aligned with the nanowire axis is increased in strength, the bulk gap may close and reopen in a topological phase transition, resulting in a topological superconductor (TSC) with MZMs at its ends. This process may be thought of as a competition between the spin-orbit coupling, the external field, and the superconducting pairing, which conspire to produce the effective $p$-wave pairing required for the Kitaev model.

The 1D Hamiltonian for the continuum nanowire model may be written in BdG form as
\bequ
\h_{\rm NW}=\left(-\frac{\hbar^2}{2m^*}\partial_x^2-i\alpha_R\partial_x\sigma_y-\mu\right)\tau_z+V_Z\sigma_x+\Delta_0\tau_x,\label{eq:NW_model}
\eequ
which acts on the Nambu spinor $\hat{\psi}_x=\left(c^\ph_{\up x},c^\ph_{\down x},c^\dagger_{\up x},-c^\dagger_{\down x}\right)^T$. In this expression, the nanowire extends along the $x$-direction, $m^*$ is the effective mass of the nanowire material, $\alpha_R$ is the Rashba spin-orbit coupling parameter, $V_Z$ is the Zeeman term, and $\Delta_0$ is the $s$-wave pairing gap induced by proximity. The Pauli matrices $\{\tau_x,\tau_y,\tau_z\}$ act in particle-hole space, while $\{\sigma_x,\sigma_y,\sigma_z\}$ act in spin space. The model is in the topological regime when the topological criterion, 
\beq
V_Z&>&\sqrt{\Delta^2+\mu^2},\label{eq:top_crit}
\eeq 
is satisfied \cite{SauNonAbelian2010,OregHelical2010,LutchynMajorana2010}. Deep in the topological phase, the Hamiltonian can be shown to reduce to that of Eq.~\eqref{eq:Kitaev} \cite{AliceaNonAbelian2011}.

The continuum nanowire model has been used as a starting point for many theoretical and numerical studies of devices believed to host MZMs (see, for example, Refs.~\cite{LiuAndreev2017,LiuConductance2019}). By incorporating components such as leads, by adjusting the dimensionality, or by adding new terms to the Hamiltonian, a variety of different experimental setups can be simulated, and the effects on the stability and behaviour of the resulting Majorana modes studied. Notably, the continuum nanowire model has been used to produce numerical conductance simulations which agree extremely well with several state-of-the-art experimental measurements \cite{LiuAndreev2017,ZhangQuantized2018}. In this way, numerical simulations have provided important supporting evidence for the existence of Majorana modes in real nanowire devices.
\subsection{Majorana Braiding\label{sec:braiding}}
Majorana fermions, being examples of non-abelian Ising anyons, are useful for quantum computing due to their ability to store information nonlocally. In the ideal case, when a pair of localised Majorana fermions are brought together, one of two things may happen: the pair may annihilate, or the pair may fuse to form an ordinary complex fermion. However, when the Majorana fermions are well separated, there is no local measurement which can be made to distinguish which outcome (or fusion channel) will result from this process. Instead, the outcome depends on the topology of the trajectories that the Majorana fermions have followed up to this point. A benefit of this nonlocal encoding of the state is that local perturbations, which are prevalent in any realistic experimental system, cannot easily affect this topology, which endows the state with an inherent robustness. On the other hand, to be useful for quantum computing, the state will need to remain coherent over timescales long enough to perform a topological operation.

In this paper, we focus on one such operation: a braid between two Majorana fermions (i.e. two sequential exchanges) which alternates between the two fusion channels. Although the braid only directly involves two Majorana modes, we require a set of four MZMs in total in order to alter the fusion channel, as the particle-hole symmetry requires the total fermion parity to be conserved. Abstractly, we can imagine creating two pairs of MZMs from the vacuum ($\{\gamma_1, \gamma_2\}$ and $\{\gamma_3, \gamma_4\}$), as shown at the top of Fig.~\ref{fig:braiding}. At this stage, if we were to bring the two MZMs from each pair back together they would annihilate. We can therefore label this state as 
\beq
\ket{n_1=0,n_2=0}=c_1c_2\ket{\Omega},
\eeq 
where $n_i=c^\dagger_ic^\ph_i$ is an occupation number, $\ket{\Omega}$ is the initial vacuum, and we have used the \emph{complex} fermion operators $c_1=\frac{1}{2}\left(\gamma_1+i\gamma_2\right)$ and $c_2=\frac{1}{2}\left(\gamma_3+i\gamma_4\right)$. [Note, however, that there is a gauge freedom in this choice]. For later use, we define this initial state as the $\ket{0}$ state of a topological qubit.

To perform the braid, we adiabatically exchange the Majorana fermions $\gamma_2$ and $\gamma_3$ twice with the same orientation, so that each one returns to its initial position but with their trajectories intertwined, as indicated in Fig.~\ref{fig:braiding}. It may be shown that a braid of this form results in the transformation $\gamma_2\to-\gamma_2$, $\gamma_3\to-\gamma_3$ \cite{AliceaNonAbelian2011}, and so
\beq
c_1&\to&c_1^\dagger\\
c_2&\to&-c_2^\dagger
\eeq
in terms of complex fermion operators. In this way, after the braid the pairs $\{\gamma_1,\gamma_2\}$ and $\{\gamma_3,\gamma_4\}$ now each fuse to form a complex fermion, and the starting state has been transformed to 
\beq
\ket{n_1=1,n_2=1}=(-)c^\dagger_1c^\dagger_2\ket{\Omega}.
\eeq
If we identify this state as the $\ket{1}$ state of a topological qubit, then the braid may be interpreted as a Pauli $\sigma_x$ operation or a NOT gate. We observe that total parity is conserved throughout this process, as required by particle-hole symmetry.

The ideal case discussed above and illustrated in Fig.~\ref{fig:braiding}(a) considers an adiabatic braid involving only four degenerate Majorana fermions. In a real system, any MZMs will necessarily be part of a much larger spectrum of states, the MZMs themselves will not be at exactly zero energy, and any braiding process will take place over a finite amount of time. All of these factors can negatively affect a braid's success. 

In this paper, we will reproduce this braid in nanowire Y-junctions and tuning forks described by the Hamiltonian in Eq.~\eqref{eq:NW_model}, following the sequence of moves shown in Fig.~\ref{fig:braiding}(b). In this setup, we set the locations of the topological regions of the wire by altering the local chemical potential $\mu$ so that the topological criterion (Eq.~\eqref{eq:top_crit}) is locally satisfied, forming MZMs at the boundary. By changing the local chemical potential as a function of time, we move the boundaries of the topological regions and, consequently, the positions of the MZMs. In contrast to the ideal case, a real nanowire has a continuum of bulk excited states: the braid must be performed slowly enough that the initial state does not mix with these excitations. On the other hand, the Majorana modes themselves will not be exactly at zero energy. Instead, the overlap between Majorana modes at different ends of the device will lead a small energy splitting, while the ideal braid assumes that the Majorana modes exist in a degenerate zero-energy subspace. Our braid must be performed fast enough that this assumption still effectively holds, and the dynamical phases that are introduced by this energy splitting must be carefully taken into account. We will study the effects of these considerations on the braiding process in the next few sections.

We note that while in our case we can study the success of a braid using numerical measures (such as wavefunction overlaps), evaluating a braid in a real device is a more challenging endeavour. Several techniques have been suggested to measure the fusion channel of a pair of Majorana fermions experimentally: In one proposal, the pair of Majoranas can be fused across a Josephson junction and the resulting Josephson current measured \cite{AliceaNonAbelian2011}. Alternatively, a Josephson junction with a valve may be used as a `parity-to-charge converter', which can either be used to detect the charge of the fusion channel directly or the channel can be measured indirectly through cyclic current measurements \cite{AasenMilestones2016,ClarkeProbability2017}.

\section{Static Properties of Majorana Y-Junctions\label{sec:static_properties}}
\subsection{System Setup\label{sec:system_setup}}
\begin{figure}
\includegraphics[scale=0.5]{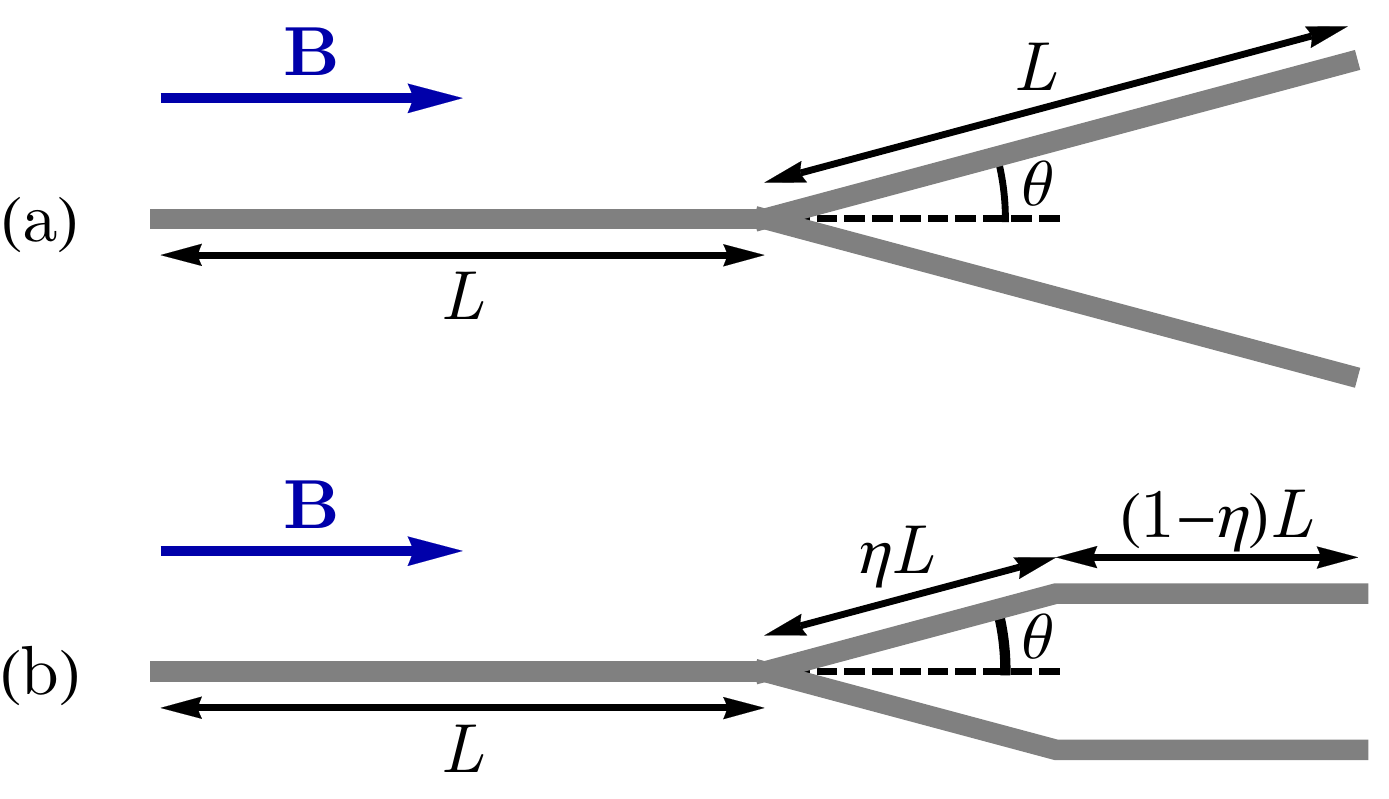}
\caption{(a) Nanowire device with a `Y-junction' geometry. Each leg has a total length of $L$ and the angle between each prong and the external magnetic field is $\theta$. (b) Nanowire device with a `tuning fork' geometry. In addition to $\theta$, the tuning fork is labelled by a parameter $\eta$, which describes the proportion of each right-hand leg that is angled before straightening out. \label{fig:ytuningfork}}
\end{figure}

To perform a braid, we require a device with at least three legs so that the Majorana fermions can remain well separated and can move around one another. For our purposes, we focus on two simple such designs: a `Y-junction' and a `tuning fork', as shown in Fig.~\ref{fig:ytuningfork}. Nanowires can be grown with either of these geometries using TASE \cite{GoothBallistic2017a}. We assume that the three legs of each device have the same length $L$, and that the prongs of the $Y$-junction each make an angle of $\theta$ with the the direction of the magnetic field, which we take to lie in the positive $x$ direction. The tuning fork design is additionally labelled by the parameter $\eta$, which gives the proportion of each right-hand leg that is angled (before becoming horizontal again). We are particularly interested in studying how the feasibility of a braid varies with these geometric parameters.

\begin{figure}
\includegraphics[scale=0.5]{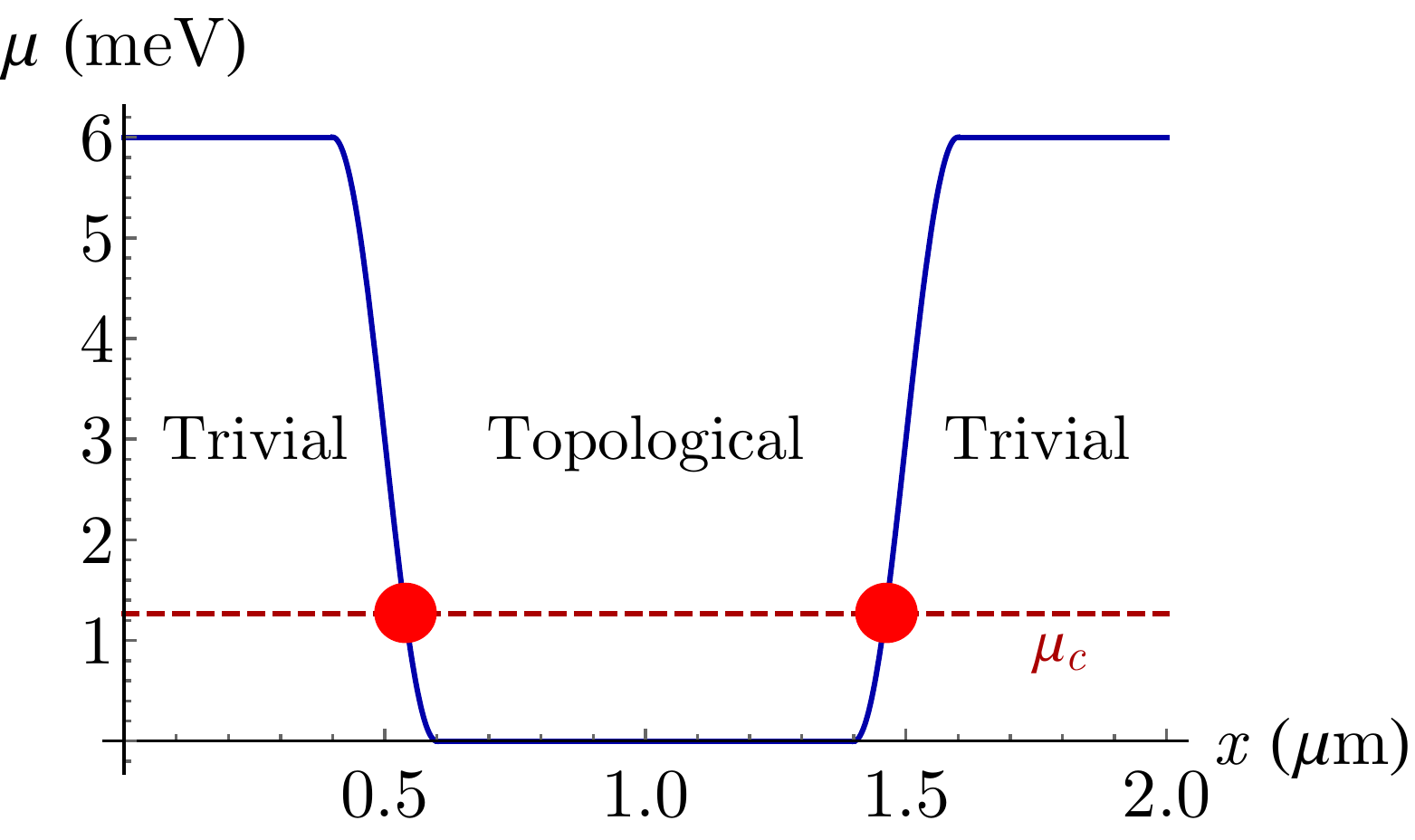}
\caption{Sample chemical potential ramping profile for (e.g.) the left leg of a nanowire Y-junction. The blue (solid) line indicates the sine-squared ramp function with two phase boundaries; the dark red (dashed) line indicates the critical value of the chemical potential at the phase transition point; the red disks schematically indicate the locations of the resulting Majorana fermions.   \label{fig:ramping}}
\end{figure}

We simulate these devices numerically by discretising the continuum nanowire model of Eq.~\eqref{eq:NW_model} on a quasi-1D lattice with lattice spacing $a$. The restriction to 1D enables us to simulate braiding processes with reasonable computational efficiently, but means that subband and orbital magnetic effects are neglected \cite{NijholtOrbital2016}. We discuss the consequences of this approximation in Sec.~\ref{sec:tuning_forks}. After discretisation, the Hamiltonian for the left leg of the Y-junction is written
\beq
\h_{0}&=&\sum_{n}\bigg\{-t(a)\bigg(\ket{n+a}\!\bra{n}+\hc\bigg)\tau_z\nonumber\\
&&-i\alpha(a)\bigg(\ket{n+a}\!\bra{n}-\hc\bigg)\sigma_y\tau_z+\Delta_0\ket{n}\!\bra{n}\tau_x\nonumber\\
&&+\bigg(-\mu+2t(a)\bigg)\ket{n}\!\bra{n}\tau_z+V_Z\ket{n}\!\bra{n}\sigma_x\bigg\},
\eeq
with
\beq
t(a)=\frac{\hbar^2}{2m^*a^2},&~~~~~~&\alpha(a)=\frac{\alpha_R}{2a},
\eeq
and where $n$ labels each site in the lattice. 

In the right-hand legs of the Y-junction, however, the magnetic field and nanowire axis are no longer parallel. The continuum Hamiltonian for a nanowire making an angle $\theta$ to the $x$-axis may be obtained from Eq.~\ref{eq:NW_model} by making the substitutions
\beq
\partial_x^2&\to&\partial_{x'}^2\equiv\partial_{x\cos\theta+y\sin\theta}^2\nonumber\\
\partial_x\sigma_y&\to&\partial_{x'}\sigma_{y'}\equiv\partial_{x'}\left[\cos\theta\sigma_y-\sin\theta\sigma_x\right],
\eeq
which is equivalent to replacing $x\to x'$ and rotating the spin-orbit coupling direction so that it remains perpendicular to the direction of propagation along the wire. In our setup, the magnetic field and spin-orbit coupling direction both lie in the $xy$-plane; however, see Refs.~\onlinecite{OscaEffects2014,RexTilting2014} for a nanowire Hamiltonian in which the magnetic field may vary in three dimensional space.

After discretisation, sections of the nanowire which make an angle of $\theta$ with the horizontal are described by the Hamiltonian
\beq
\h_{\theta}&=&\sum_{n'}\bigg\{-t(a)\bigg(\ket{n'+a}\!\bra{n'}+\hc\bigg)\tau_z\nonumber\\
&&-i\alpha(a)\bigg(\ket{n'+a}\!\bra{n'}-\hc\bigg)\left(\cos\theta\sigma_y-\sin\theta\sigma_x\right)\tau_z\nonumber\\
&&+\Delta_0\ket{n'}\!\bra{n'}\tau_x+\bigg(-\mu+2t(a)\bigg)\ket{n'}\!\bra{n'}\tau_z\nonumber\\
&&+V_Z\ket{n'}\!\bra{n'}\sigma_x\bigg\}.\label{eq:ham_theta}
\eeq
We will find that the misalignment of the field and the nanowire leg has a significant effect on the eigenstates of the system. In addition to the above transformations, we must also treat the central site of the Y-junction with care: since it connects to all three legs, we replace the diagonal discretisation offset term $2t(a)\tau_z$ with $3t(a)\tau_z$, so that the model has a well-defined continuum limit as $a\to0$.

We build the discretised system and perform exact diagonalisation using the KWANT python package \cite{GrothKwant2014}, taking the lattice spacing $a$ small enough to avoid discrete artefacts. The parameters we choose are inspired by existing numerical works, based on InSb nanowires and aluminium superconductors, which have been shown to agree extremely well with experimental data \cite{LiuAndreev2017,ZhangQuantized2018}. We also take additional input from the device geometries and materials used in the TASE process \cite{GoothBallistic2017a}. Our starting parameters (which we take as a `best case' for such devices) are $m^*=0.015~m_e$, $\alpha_R=0.5~{\rm eV~\AA}$, $L=2~\mu m$, $\Delta_0=0.8~{\rm meV}$, and $V_Z=1.5~{\rm meV}$. This value of $V_Z$ corresponds to a magnetic field strength of $B\approx1.3$~T, assuming $V_Z=\frac{1}{2}g_{\rm eff}\mu_B B$ with $g_{\rm eff}\approx40$ \cite{LiuAndreev2017,ZhangQuantized2018}. We choose a lattice spacing of $a=10^{-8}~{\rm m}$ (so that the device consists of 600 sites), which we find is small enough that the low energy band structure and dynamical properties of the system are stable and immune from discreteness effects. These best-case parameters will be used throughout the simulations in Secs.~\ref{sec:static_properties} and \ref{sec:braiding_simulations}.

We note that our choices for $L$ and $\Delta_0$ are slightly larger than what has been assumed in previous simulations, but are values we believe are achievable using the TASE technique and using nitride-based superconductors \cite{LutchynMajorana2018}; we study the effect of reducing these parameters in Sec.~\ref{sec:tuning_forks}. We change between the topological and trivial superconducting regimes by adjusting the chemical potential, choosing $\mu_{\rm top}=0$ and $\mu_{\rm triv}=6~{\rm meV}$, where the topological phase transition occurs at $\mu_c=1.279~{\rm meV}$ for the horizontal wire. The parameters $\theta$ and $\eta$, which define the geometry of the devices, will be varied. 

As described in Sec.~\ref{sec:braiding}, a braid can be performed by moving the topological regions of the wire (and consequently, the Majorana fermions at their boundaries) through the sequence shown in Fig.~\ref{fig:braiding}(b). Numerically, this is achieved by changing the local chemical potential $\mu(x,t)$ as a function of position and time. However, there is considerable freedom in deciding the steepness and functional form of the chemical potential at the boundary between different phases, where it varies between $\mu_{\rm top}$ and $\mu_{\rm triv}$. We use a sine-squared ramping potential based on that of Ref.~\cite{SekaniaBraiding2017}, which takes the form
\beq
\mu(x,t)&=&\mu_{\rm triv}\sin^2\left[\frac{\pi}{2}r\left[\beta \left(x- x_c(t)\right)\right]\right],\label{eq:ramp_function}
\eeq
where $r(x)$ is the linear ramp function
\beq 
r(x)&=&\left\{\begin{array}{cc}
0 & x<0 \\
x & 0\leq x \leq 1\\
1 & x>1
\end{array}\right..\label{eq:linear_ramp}
\eeq
In this expression, $\beta=5~(\mu{\rm m})^{-1}$ sets the steepness of the ramp (chosen to minimise the numerical Majorana wavefunction size), and $x_c(t)$ identifies the position of the base of the ramp as a function of time. In Ref.~\cite{SekaniaBraiding2017} the sine-squared ramp was found to yield better braiding results than the bare linear ramp, and it should also give a better approximation to the smooth chemical potential profiles that would be realised using side gates in an experiment. In Appendix~\ref{app:gates}, we demonstrate that a ramp potential similar to this may be generated using a realistic arrangement of side gates. There, we estimate that approximately twenty gates would be required on each leg of the nanowire device to reproduce a ramping potential to within about 5\%.

An illustration of two phase boundaries described by the sine-squared ramping profile is shown in Fig.~\ref{fig:ramping}. During the braiding process, these phase boundaries are moved throughout the device so that the two Majoranas attached to them form a braid. [The other two Majorana fermions remain at the ends of the nanowire legs throughout the process].

\subsection{Low-lying Bulk States}
\begin{figure}[t]
\includegraphics[scale=0.3]{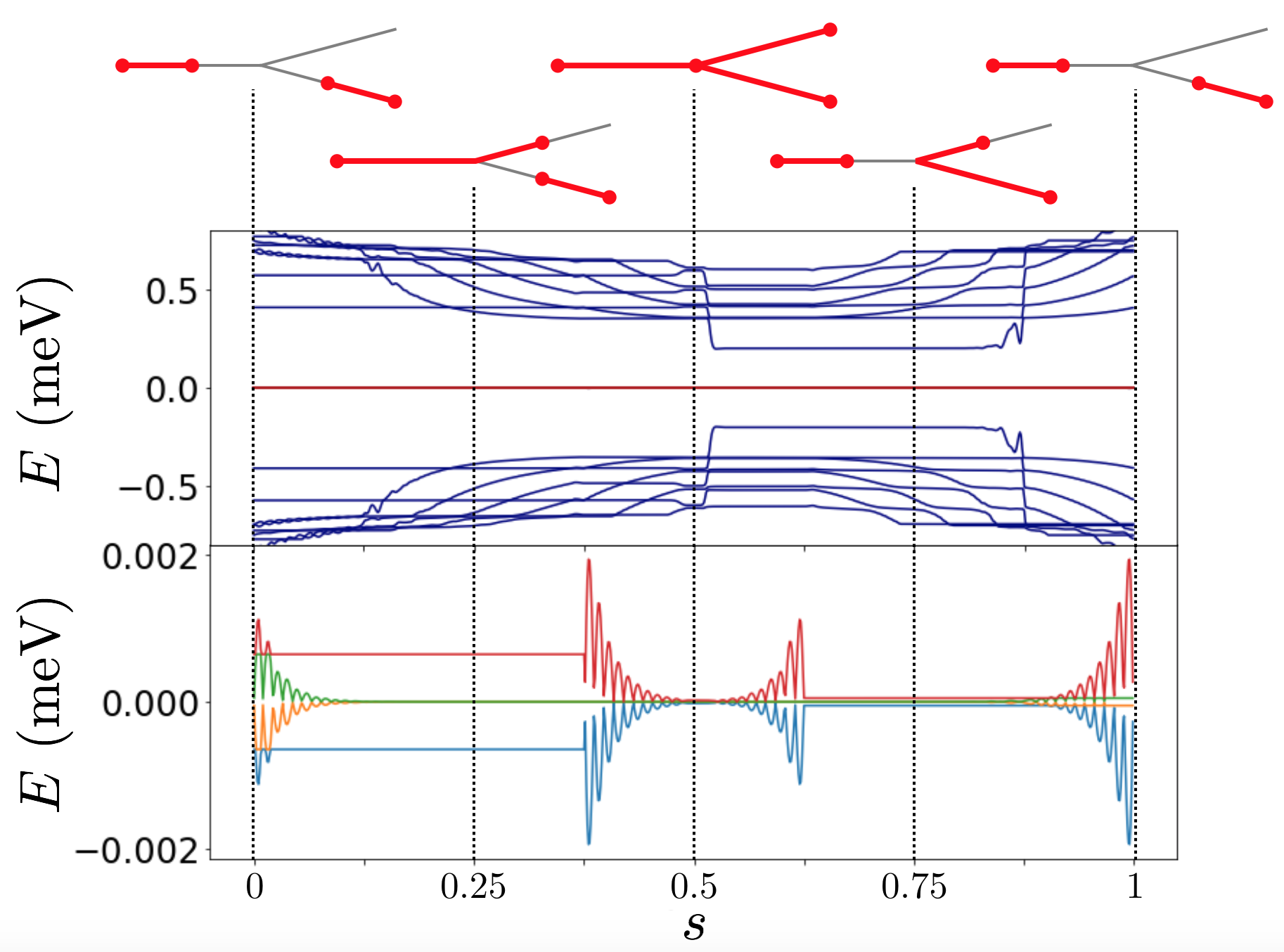}
\caption{The twenty smallest-magnitude instantaneous energy levels for a Majorana Y-junction with $\theta=15^\circ$, as a function of `time' $s$ during the exchange described in the main text. Note that this exchange corresponds to \emph{half} a complete braid, and that we have labelled the progress of the exchange operation with the label $0\leq s\leq 1$. Since this plot shows the energy levels of the \emph{instantaneous} Hamiltonian, the label $s$ is not associated with any physical time scale. \emph{Top}: The instantaneous position of the four Majorana modes at five points during the exchange. \emph{Middle}: The twenty smallest-magnitude energy levels as a function of $s$ during the exchange. Energy levels corresponding to Majorana modes are shown in red (close to zero), while energy levels corresponding to bulk states are shown in blue (above and below). \emph{Bottom}: A zoomed-in plot of the four Majorana energy levels as a function of $s$ during the exchange. \label{fig:braid_bands}}
\end{figure}
\begin{figure*}
\includegraphics[scale=0.45]{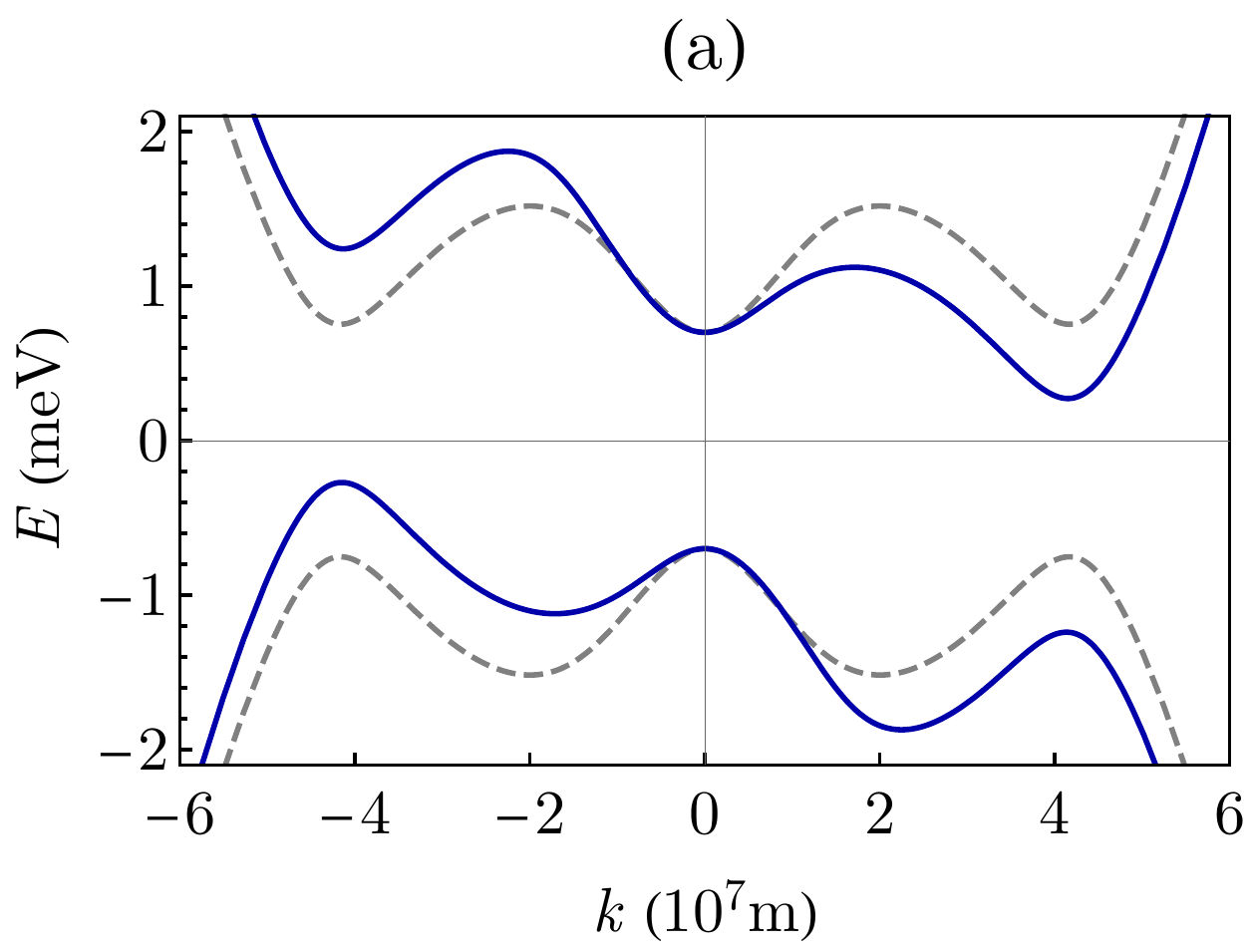}\hspace{2mm}
\includegraphics[scale=0.45]{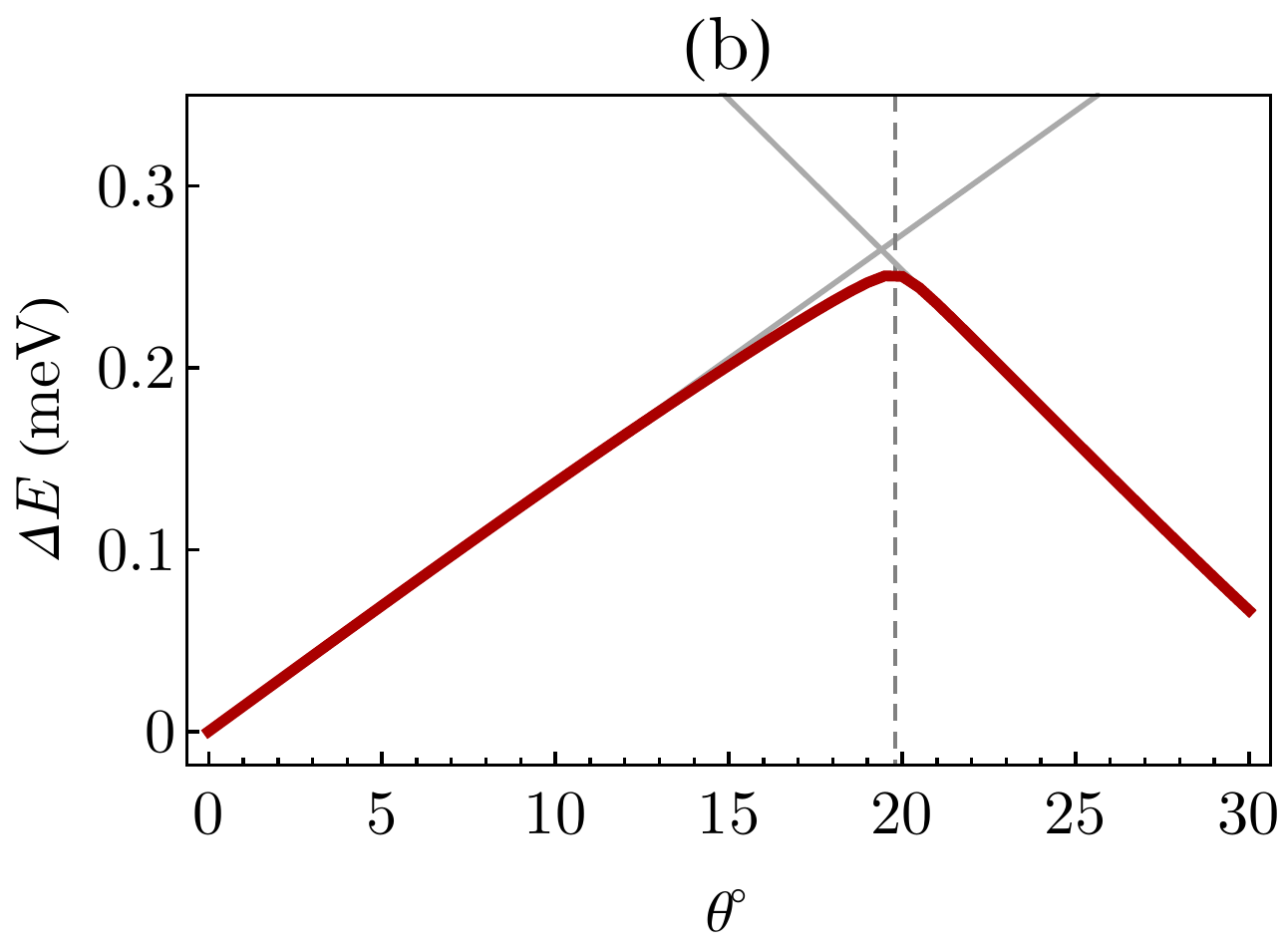}\hspace{2mm}
\includegraphics[scale=0.45]{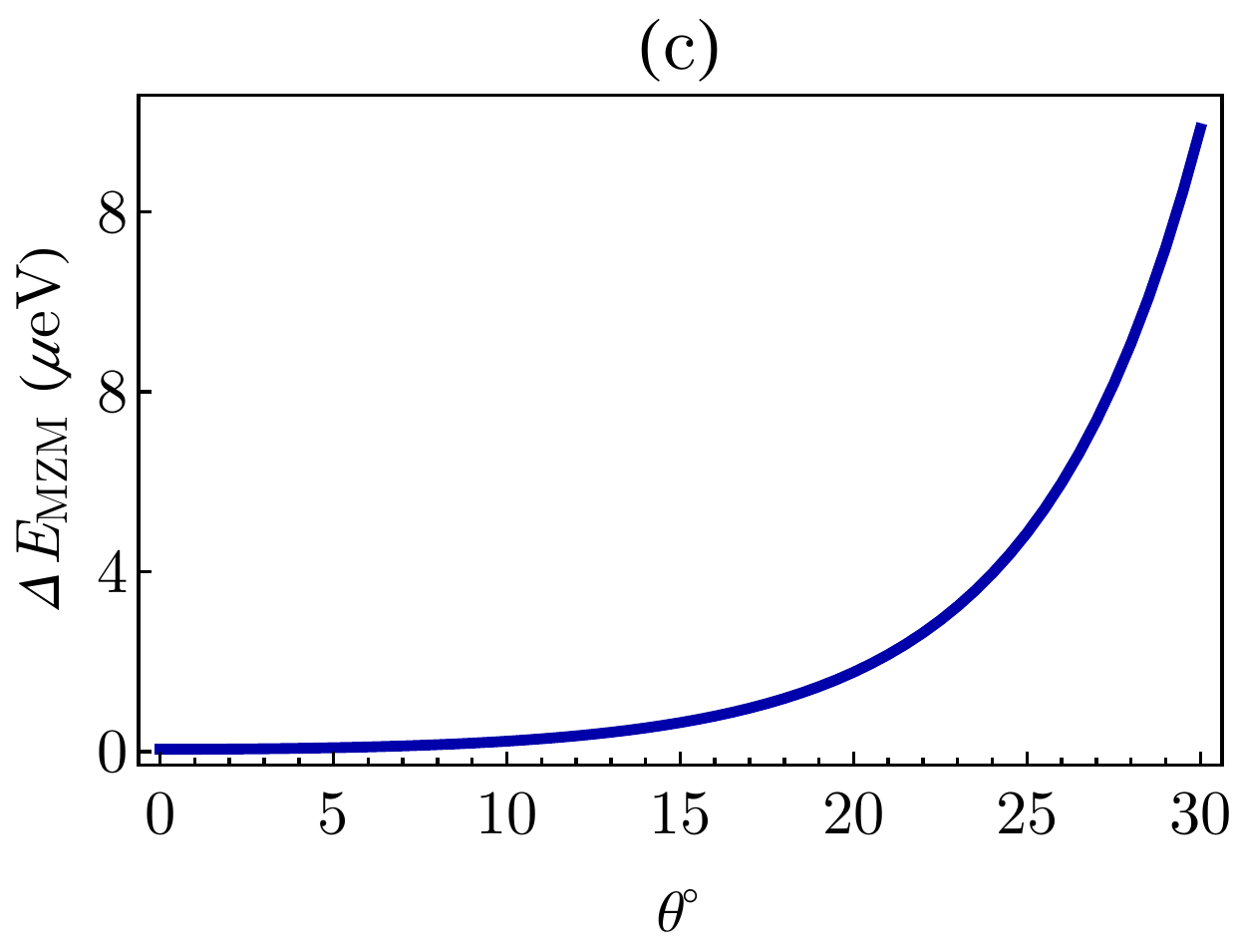}
\caption{(a) Low-lying bands of the continuum version of Hamiltonian~\eqref{eq:ham_theta} for $\theta=0^\circ$ (grey dashed line) and $\theta=20^\circ$ (blue solid line). The band structure becomes asymmetric and the (indirect) gap gets closer to zero as $\theta$ is increased. (b) Numerical bulk gap above zero for the Y-junction device as a function of half-angle $\theta$ [red (dark gray) solid line]. On the left, the gap is dominated by incipient Majorana modes forming at the central junction, while on the right the gap is dominated by the bulk gap collapse. Light gray solid lines show linear fits in the low-$\theta$ and high-$\theta$ regions. Vertical gray dashed line indicates the optimum value of $\theta$ which maximises the bulk gap. (c) Largest energy splitting of the Majorana modes (measured relative to zero at $s=0.125$) as a function of $\theta$. The splitting increases approximately exponentially with $\theta$. \label{fig:numerical_energies}
}
\end{figure*}
For a successful braid, we require (at the very least) that the quasi-degenerate Majorana subspace remains separated from other excited states by an energy gap $\Delta E$. The size of this energy gap will impose limits on the speed at which we can perform the braid without causing diabatic excitations. For this reason, we first study the instantaneous low-lying energy levels near the Majorana subspace as the braid progresses. An example of this is shown in Fig.~\ref{fig:braid_bands} for the Y-junction with parameters as given above, and with $\theta=15^\circ$. In this plot, the progress of the exchange operation is labelled by the parameter $0\leq s\leq1$, which may loosely be thought of as a dimensionless time parameter. [Note, however, that the energy levels plotted are of the instantaneous Hamiltonian, and so $s$ is not associated with any physical time scale].

We see from Fig.~\ref{fig:braid_bands} that, as hoped, the instantaneous Majorana energy levels remain well-separated from the bulk states throughout the braid. However, there appear to be two distinct types of low-lying excitation: first, there is a broad spectrum of `general' bulk states, and secondly there is a single pair of low-lying states which approach the Majorana levels for $0.5\lesssim s\lesssim 0.9$.

The broad spectrum corresponds to bulk states that are also present in a system with periodic boundary conditions, and which numerically are found to be delocalised across the device. These define the bulk energy gap, which is a (complicated) function of the parameters of Hamiltonian~\eqref{eq:ham_theta}. Importantly, this bulk gap decreases when $\theta$ increases, as the external magnetic field and the effective spin-orbit coupling field (which is perpendicular to $\bB$ when $\theta=0^\circ$) start to become aligned. This effect is apparent in the band structure of a single straight nanowire, shown for a continuous periodic system in Fig.~\ref{fig:numerical_energies}(a): at $\theta=0^\circ$ the electron and hole bands are gapped and symmetric about $k=0$, but when $\theta=20^\circ$, the bands become tilted and the bulk gap reduces. This is associated with the breaking of the symmetry $\sigma_x\h_\theta(\bk)\sigma_x=\h_\theta(-\bk)$ of Hamiltonian~\eqref{eq:ham_theta} when $\theta\neq0$ [see Ref.~\cite{NijholtOrbital2016} for further details of symmetries in TSC nanowire models]. We find numerically (for both the periodic system and the Y-junction device) that the bulk band gap collapses approximately linearly with $\theta$, from its bare value near $\Delta_0$ at $\theta=0^\circ$, to zero at $\theta\approx35^\circ$, as shown in Fig.~\ref{fig:numerical_energies} (b). Indeed, Refs.~\onlinecite{OscaEffects2014,RexTilting2014} consider this band tilting for the infinite continuum nanowire model, and find a bulk gap closure that is consistent with our numerical results for the Y-junction device. In particular, Ref.~\onlinecite{RexTilting2014} demonstrates that the critical angle should satisfy (in our notation)
\beq
\theta_c=\mathrm{arcsin}\left[\Delta_0/V_Z\right]\approx32.2^\circ,
\eeq
which agrees very well with our numerical results.

The second type of low-energy modes, the single pair of states which emerges from the bulk during the evolution, is specific to the three-legged geometry: numerically, these states have most of their density concentrated at the junction between the three legs. In the period where these modes are most significant (between $0.5\lesssim s\lesssim 0.9$ in Fig.~\ref{fig:braid_bands}), the two right-hand legs are in the topological regime and may be interpreted as two separate TSC nanowires forming a junction. When two such nanowires (each hosting a pair of MZMs) are brought together, the two MZMs that meet at the junction couple through the (approximate) Hamiltonian term
\beq
\h_{12}&\propto&-i\Gamma\cos\left(\frac{\phi_A-\phi_B}{2}\right)\gamma_1\gamma_2,\label{eq:majorana_coupling}
\eeq
where $\gamma_{1,2}$ are the relevant Majorana operators, $\Gamma$ is a coupling strength, and $\phi_A$ and $\phi_B$ are the effective superconducting pairing phases in each nanowire \cite{KitaevUnpaired2001,AliceaNonAbelian2011}. In general, this means that when two TSC nanowires are brought together, the two MZMs at the junction are gapped out, leaving a single, longer TSC nanowire with a single pair of MZMs. However, in the fine-tuned case where $\phi_A-\phi_B=\pi$ (a `$\pi$-junction' \cite{AliceaNonAbelian2011}), the coupling term vanishes and the resulting system has \emph{four} Majorana zero modes.

For the realistic nanowire model we are using, the induced superconducting pairing depends on the spin-orbit coupling direction \cite{OregHelical2010,LutchynMajorana2010,AliceaNonAbelian2011}, which is different in each leg. With a half-angle of $\theta$, the effective superconducting pairing phases are proportional to $e^{i\phi_{\rm UR}}\propto e^{-i\theta}$ and $e^{i\phi_{\rm LR}}\propto e^{-i\left(\pi-\theta\right)}$ (where $\phi_{\rm UR}$ and $\phi_{\rm LR}$ refer to the upper-right and lower-right legs, respectively). This means that as $\theta$ decreases, the two legs get closer and closer to forming a $\pi$-junction: the single pair of low-lying states correspond to these \emph{incipient} Majorana modes that would arise exactly at $\theta=0^\circ$. The resulting gap is therefore greater at \emph{larger} values of $\theta$. Expanding Eq.~\eqref{eq:majorana_coupling}, we expect the energy gap caused by incipient Majorana modes to increase linearly with $\theta$ (or as $\sin\theta$ for large enough values of $\theta$).

These two sources of low-lying modes, the bulk gap closure and the incipient Majoranas, become dominant at different values of $\theta$. The optimum (largest) energy gap above the MZM subspace will be obtained for a half-angle which is greater than zero (to avoid a $\pi$-junction), but not so great that the bulk gap closure starts to become limiting. For our base parameter values, we find that this is achieved for a critical value $\theta_c\approx20^\circ$, as demonstrated in Fig.~\ref{fig:numerical_energies}(b). Note, however, that the energy gap at this angle is lower than its value for a single straight nanowire at $\theta=0^\circ$. In Sec.~\ref{sec:tuning_forks}, we study how this angle dependence changes for the tuning fork geometry and with different parameter values.

In addition to a large energy gap to excitations, we also desire the modes corresponding to Majorana fermions to be as close to zero energy as possible: in the ideal case, all four MZM states would be exactly degenerate. In a real system, however, the the Majorana zero modes are extended over some localisation length $\xi$, and modes from different regions of the device overlap in space, leading to an energy splitting. For two modes separated in space by a distance $\ell$, the splitting is proportional to $\Delta E_{\rm MZM} \propto e^{-\ell/\xi}$. In contrast to the ideal case, this energy splitting means that our braid operation will need to be performed fast enough that the different MZMs can be considered as a single subspace (i.e. the braid should be able to mix states freely within this subspace). Even then, the splitting will generally introduce different dynamical phases for different states in the subspace, which will need to be taken into account.

We can see the energy splitting between different Majorana states in the lower panel of Fig.~\ref{fig:braid_bands}:  for example, there are two modes in red and blue with approximate energy levels $E\approx\pm0.65~\mu{\rm eV}$ that remain constant for $0\lesssim s\lesssim 0.4$. These correspond to the two MZMs initially localised on the lower-right leg of the Y-junction, which remain stationary throughout this part of the braid. They have a larger energy splitting than the other MZM modes because they are closer together and because they exist in the region of the wire with a smaller bulk gap (since $\theta=15^\circ$ here). We also observe oscillations whenever two Majorana modes get closer together or move further apart.

Since the Majorana energy splitting will be important in interpreting the braiding results of the next section, we plot the energy of the largest Majorana mode (relative to zero) at $s=0.125$ as a function of $\theta$ in Fig.~\ref{fig:numerical_energies}(c). We choose this specific gap in the low-energy subspace, as it is the largest constant gap and persists long enough to have noticeable effects. We expect the oscillatory splitting, on the other hand, to generate dynamical phases that on average cancel out. 

We note that the energy splitting shown in Fig.~\ref{fig:numerical_energies}(c) increases approximately exponentially with $\theta$ and does not show any particular behaviour at the critical angle of $\theta_c\approx20$. This is consistent with the fact that the localisation length of the MZM is inversely proportional to the (local) bulk pairing gap \cite{AliceaNew2012}. In this case, we found numerically [as in Fig.~\ref{fig:numerical_energies}(b)] that the bulk gap decays approximately linearly as $\Delta E_{\rm bulk}\propto\Delta_0-c\theta$ for some constant $c$. This implies a MZM splitting proportional to $\Delta E_{\rm MZM} \propto e^{-\kappa\left[\Delta_0-c\theta\right]}$ (for some approximately constant $\kappa$), which reproduces the behaviour of Fig.~\ref{fig:numerical_energies}(c). 

\section{Braiding Simulations in Majorana Y-Junctions\label{sec:braiding_simulations}}
\subsection{Braiding Numerics}
\begin{figure*}[t]
\includegraphics[scale=0.39]{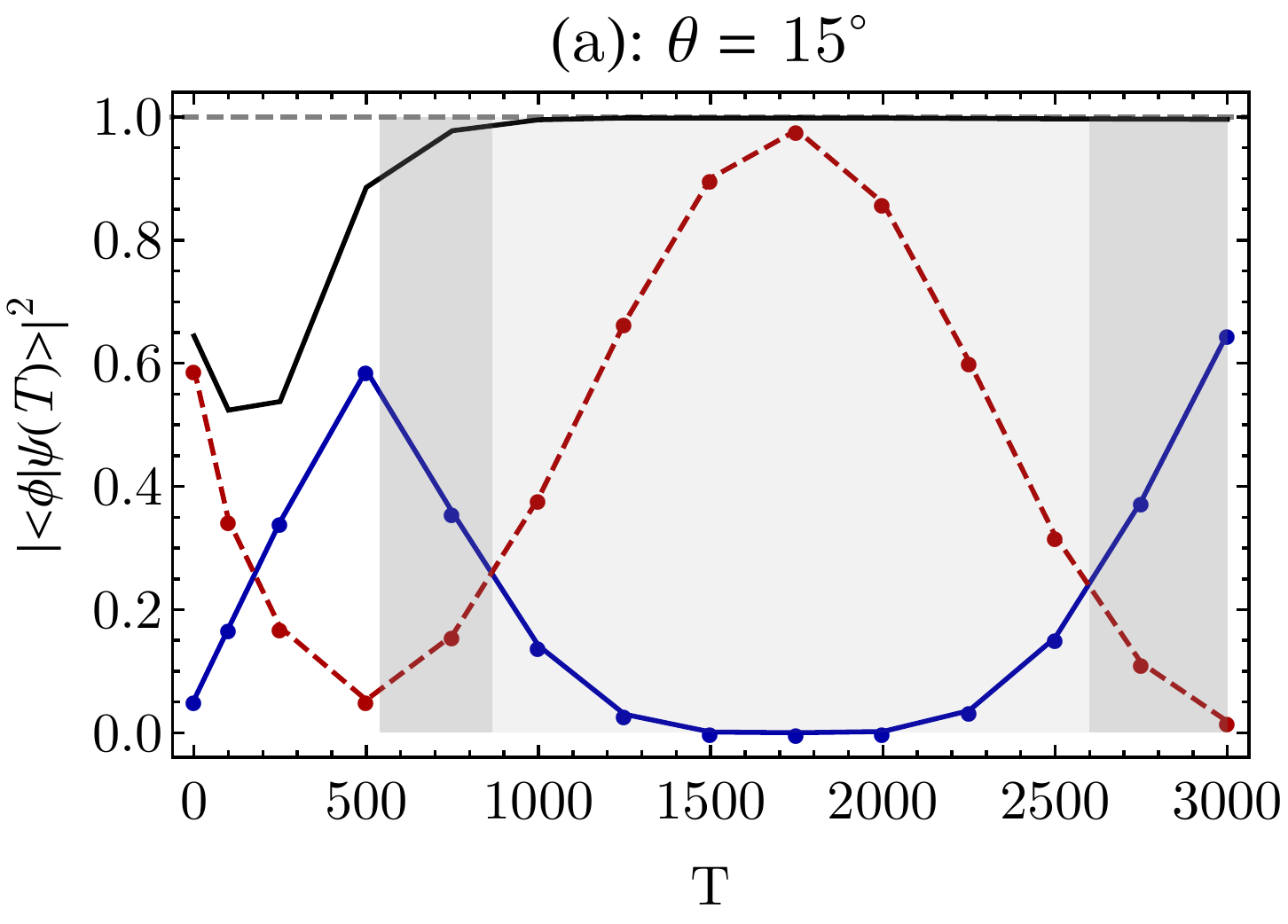}\hspace{1mm}
\includegraphics[scale=0.39]{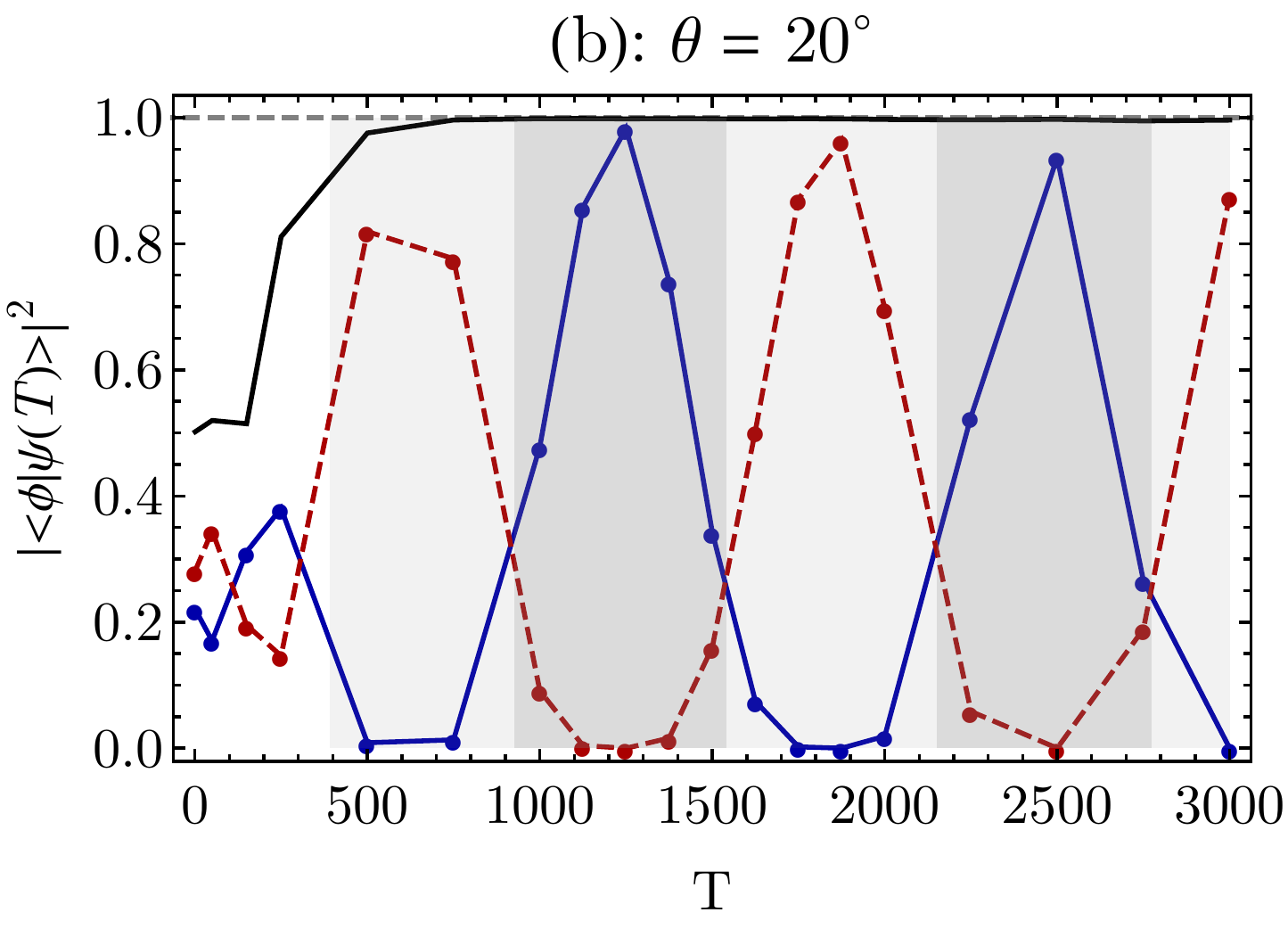}\hspace{1mm}
\includegraphics[scale=0.39]{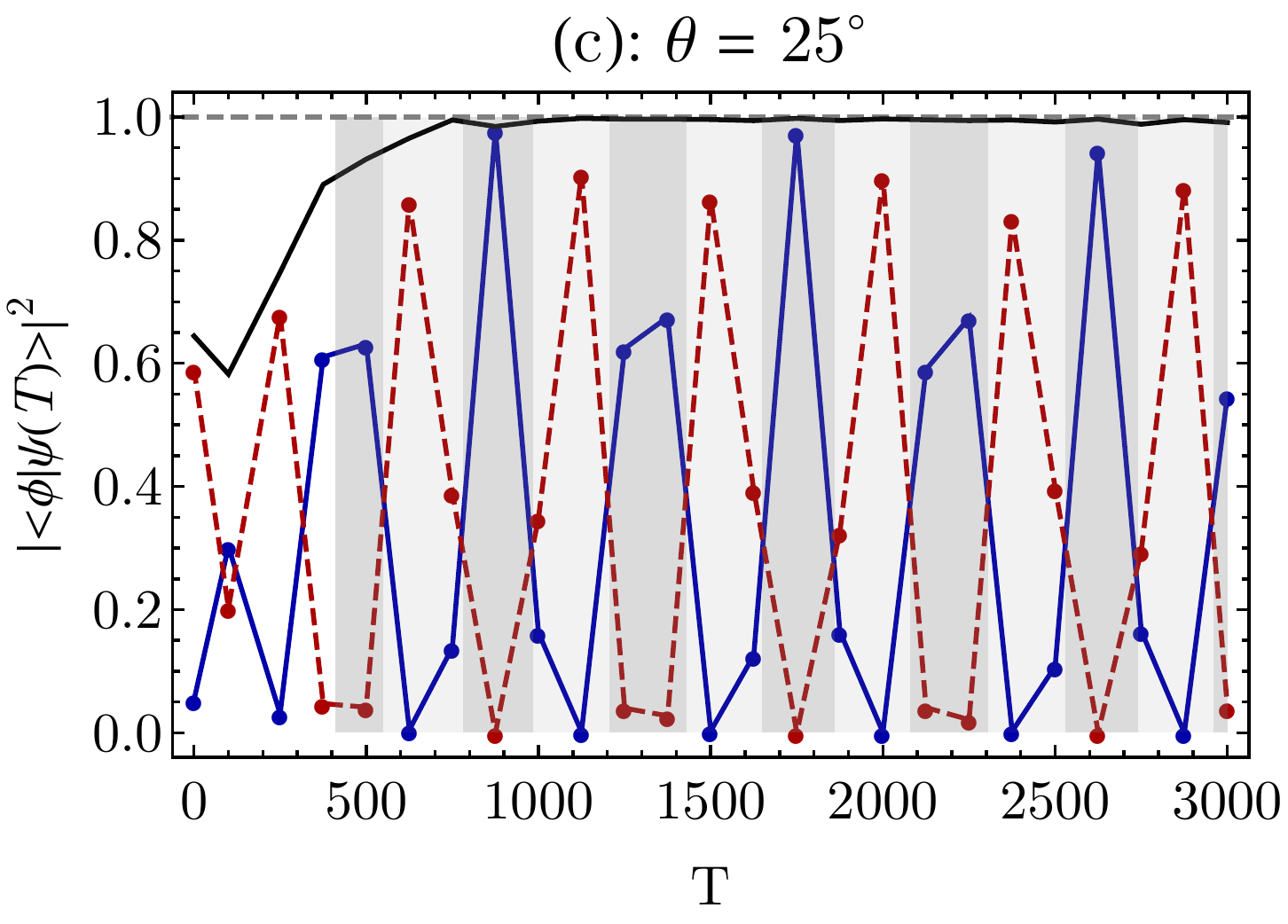}
\caption{Square overlaps of final state with initial states at the end of a braid for (a) $\theta=15^\circ$, (b) $\theta=20^\circ$ and (c) $\theta=25^\circ$. Blue solid lines (with points) indicate the overlap with $\ket{\phi}=\ket{1_+}$, the ideal final state. Red dashed lines (with points) indicate the overlap with $\ket{\phi}=\ket{1_-}$, the starting state. Black solid lines (without points) indicate the total square overlap with all four states within the initial zero-energy subspace. The dark gray shading indicates regions where the overlap with the ideal final state $\ket{1+}$ is greater than the overlap with the initial state $\ket{1-}$, and roughly indicate ranges of $T$ for which the braid is successful. The lighter grey shading (including the dark grey regions) indicates values of $T$ for which the overlap with the total Majorana subspace is greater than 95\%, and the braiding process can be interpreted as adiabatic with respect to the excited states. See main text for details. We recall that one unit of time in our units corresponds to approximately 4.14~ps in physical units. \label{fig:braiding_simulations}}
\end{figure*}

We now see how these instantaneous energy considerations play out when we simulate a finite-time braiding operation. As discussed above, we perform a braid numerically by sweeping the chemical potential as a function of space and time, following the ramp function defined in Eq.~\eqref{eq:ramp_function} and the protocol illustrated in Fig.~\ref{fig:braiding}(b). We will assume that the braid takes place over a time scale $0\leq t\leq T$, where the total braiding time $T$ will be varied. In our units (where the Hamiltonian parameters are given in meV), one unit of time corresponds to $4.14$~ps.

In the ideal case, we would calculate the time-evolution operator at a given time $t$ through the relation
\beq
U(t)&=&\mathcal{T}\exp\left[-i\int_0^tH(t')\,\dd t'\right],
\eeq
where $\mathcal{T}$ indicates time ordering and where $H(t)$ is the time-dependent Hamiltonian that incorporates the changes to the chemical potential $\mu(x,t)$. Numerically, however, we must split the time-evolution operator into a finite number of steps ($N$) and use the relation
\beq
U(t)&=&\lim_{N\to\infty}\prod_{j=1}^{j=N}e^{-iH(j\Delta t)},
\eeq
where $\Delta t=t/N$, to obtain a discrete approximation to $U(t)$. In practice, we increase $N$ until the expression for $U(t)$ converges for a given time $t$. To calculate the factors $e^{-iH(j\Delta t)}$ efficiently, we use a Chebychev expansion which converges rapidly with just a small number of terms \cite{Tal-Ezeraccurate1984}. We vary the chemical potential $\mu(x,t)$ by changing the offset of the ramp function $x_c(t)$ linearly with time during the braiding process [see Eq.~\eqref{eq:ramp_function} and Fig.~\ref{fig:braiding}(b)].

To perform a braid in an interacting system, we would take the many-body ground state (consisting of a Slater determinant of all occupied single-particle states including MZMs) and act on it with the many-body unitary evolution $U^{\rm mb}_{\rm braid}(T)$, which is a braid that takes place over a time period $T$. For our noninteracting system, we will instead consider the simpler action of the single-particle unitary evolution $U_{\rm braid}(T)$ on a single-particle state that acts as a proxy for the many-body ground state, following the method of Ref.~\onlinecite{AmorimMajorana2015}. In the absence of any interacting terms in the Hamiltonian, this procedure provides an equivalent probe of the non-abelian statistics of the MZMs.

Specifically, we recall that at $t=0$ there are four states close to zero energy that correspond to the four MZMs we expect to form under the starting arrangement of the chemical potential. The two states closest to zero energy (corresponding to orange and green in the lower panel of Fig.~\ref{fig:braid_bands}) are equal superpositions of the two Majorana fermions on the left leg of the Y-junction (labelled $\gamma_1$ and $\gamma_2$ in Fig.~\ref{fig:braiding}(b)). These states are equal superpositions of $\gamma_1$ and $\gamma_2$ due to the splitting that arises from their nonzero overlap. We write the negative/positive energy states as $\ket{1_-}$ / $\ket{1_+}$ and associate them with the (complex fermion) operators $c_1^\ph=\frac{1}{2}\left(\gamma_1+i\gamma_2\right)$ and $c_1^\dagger=\frac{1}{2}\left(\gamma_1-i\gamma_2\right)$ respectively, as in the discussion in Sec.~\ref{sec:braiding}. Similarly, we associate the other two states close to zero energy (blue and red in the lower panel of Fig.~\ref{fig:braid_bands}) with operators $c_2^\ph=\frac{1}{2}\left(\gamma_3+i\gamma_4\right)$ and $c_2^\dagger=\frac{1}{2}\left(\gamma_3-i\gamma_4\right)$ and write the corresponding states as $\ket{2_-}$ and $\ket{2_+}$. These states are associated with Majorana fermions $\gamma_3$ and $\gamma_4$, which are initially localised on the lower-right leg of the Y-junction. Note that these states are defined only for the chemical potential conformation at $t=0$.

As discussed in Sec.~\ref{sec:braiding}, the ideal braiding operation should map operators $c_1\to c_1^\dagger$ and $c_2\to -c_2^\dagger$ (up to a choice of signs). In terms of states, this would map $\ket{1_-}\to\ket{1_+}$ and $\ket{2_-}\to-\ket{2_+}$ (up to an overall phase). We can therefore test the success of the braid by starting at $t=0$ with the state
\beq
\ket{\psi(0)}&=&\ket{1_-}
\eeq
and acting on this with the unitary evolution operator to obtain
\beq
\ket{\psi(t)}&=&U_{\rm braid}(t)\ket{\psi(0)}.
\eeq
At the end of the evolution we compare the overlap of the state $\ket{\psi(T)}$ with the initial states $\ket{1_-}$ and $\ket{1_+}$. If the braid is successful, we should find
\beq
\left|\big\langle1_-\big|\psi(T)\big\rangle\right|^2&=&0\nonumber\\
\left|\big\langle1_+\big|\psi(T)\big\rangle\right|^2&=&1,\label{eq:ideal_overlaps}
\eeq
indicating that all the weight from state $\ket{1_-}$ has been transferred to state $\ket{1_+}$. In general, however, the overlaps will not take these ideal values, and there may be additional nonzero overlaps between $\ket{\psi(T)}$ and other states (either excited states or other states from the MZM subspace).
\subsection{Braiding Results\label{sec:braiding_results}}
If we perform the braid adiabatically (by numerically projecting onto the subspace of states close to zero energy at each instant of the evolution), then we obtain wavefunction overlaps that are exactly as in Eq.~\eqref{eq:ideal_overlaps}: in this ideal case, we have assumed that the braiding time $T\to\infty$ and that the Majorana states are degenerate, and the braiding operation reproduces the ideal theoretical prediction. In the more realistic case, however, where we perform the evolution unitarily, the gap to excited states and energy splitting within the Majorana subspace leads to deviations from this ideal behaviour. In Fig.~\ref{fig:braiding_simulations}, we plot the square overlaps $\left|\big\langle\phi\big|\psi(T)\big\rangle\right|^2$ of the final state $\ket{\psi(T)}$ with a variety of initial states $\ket{\phi}$, for $\theta=15^\circ$, $20^\circ$ and $25^\circ$. We recall that from considerations of the instantaneous energy gaps, $\theta=20^\circ$ is close to the optimal case (with maximum energy gap to excitations), while smaller or larger angles lead to a suppression of the gap due to incipient Majorana modes or bulk gap closure, respectively.

Fig.~\ref{fig:braiding_simulations} plots the square overlap with the ideal final state $\ket{1_+}$ (in solid blue), the square overlap with the starting state $\ket{1_-}$ (in dashed red), and the total square overlap with the initial Majorana subspace (in solid black), for each value of $\theta$. These overlaps are plotted as a function of the total braiding time $T$ and are the overlaps that would be measured immediately after the braiding protocol has been completed. Unlike in the adiabatic case, the probability of the braid being successful (given by $\left|\big\langle1_+\big|\psi(T)\big\rangle\right|^2$) oscillates as a function of the total braiding time $T$, and is suppressed significantly if the braid is performed too quickly. The dark grey shading indicates ranges of $T$ for which the braiding may approximately be viewed as `successful', in which the overlap with ideal final state $\ket{1+}$ is larger than the overlap with the initial state $\ket{1-}$. The lighter grey shading (including the dark grey areas) indicates values of $T$ for which the process may approximately be viewed as adiabatic with respect to the excited states, for which the overlap with the total Majorana subspace is greater than 95\%.

\begin{figure*}[t]
\includegraphics[scale=0.42]{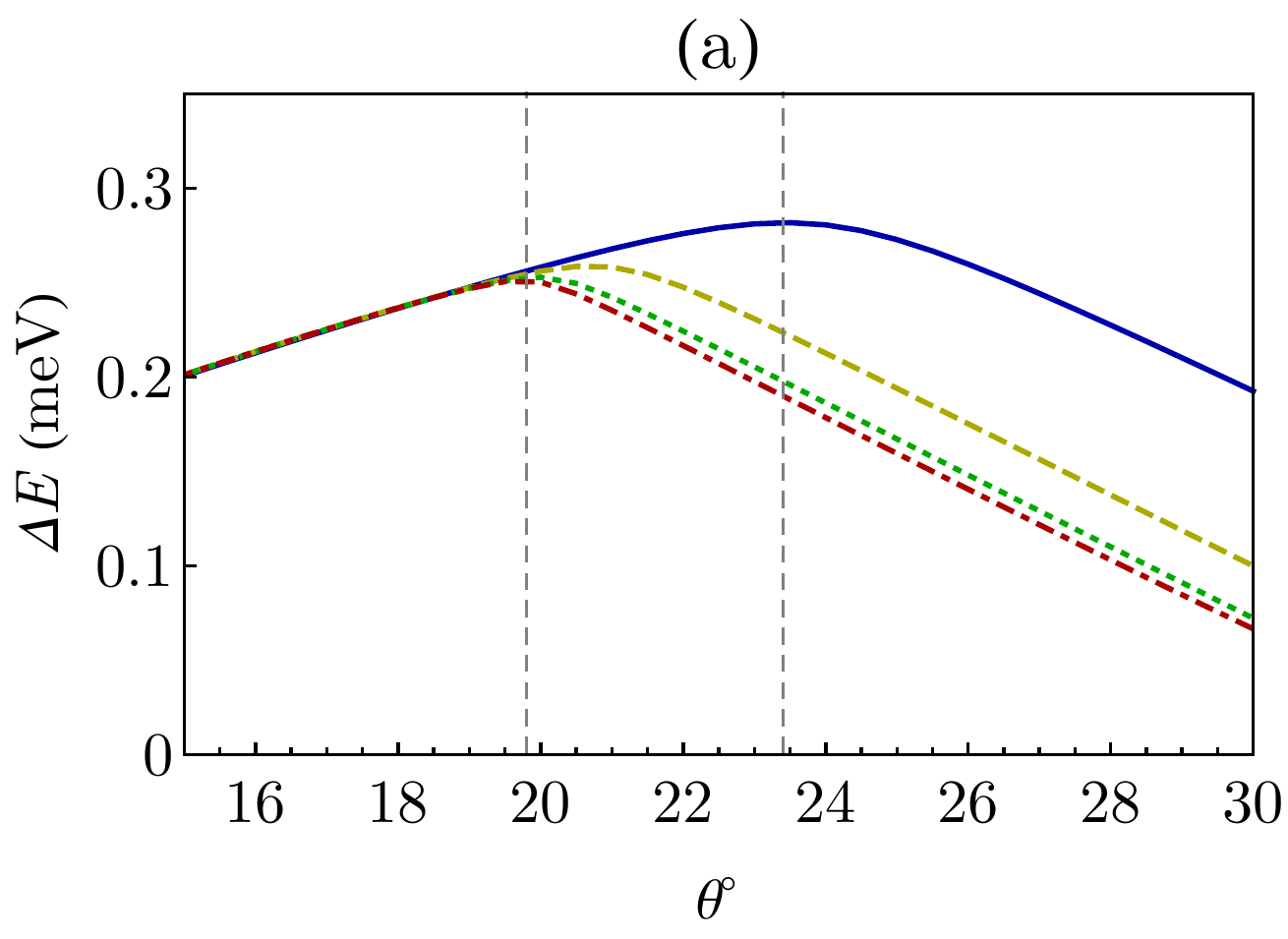}\hspace{1mm}
\includegraphics[scale=0.42]{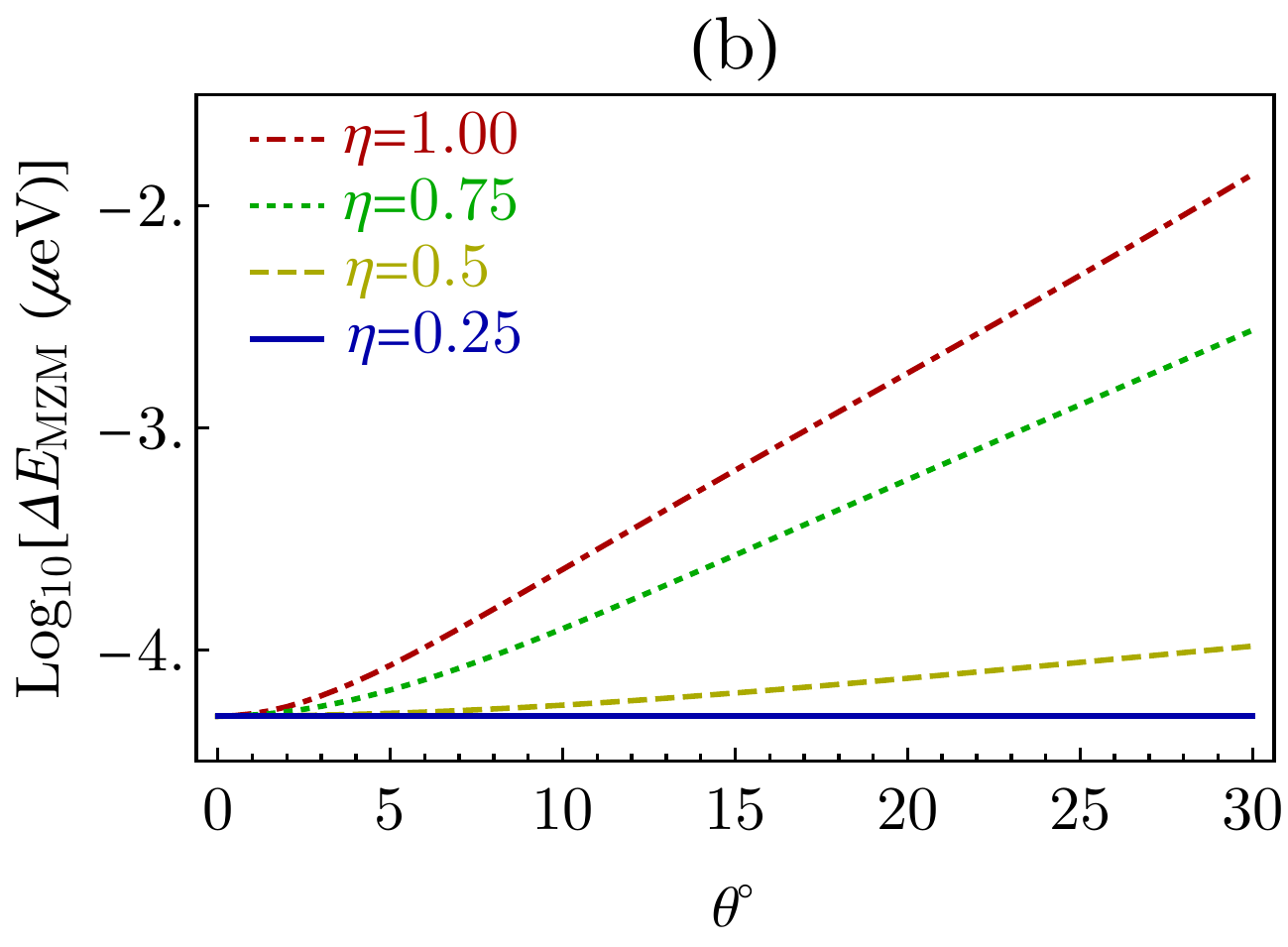}\hspace{1mm}
\includegraphics[scale=0.425]{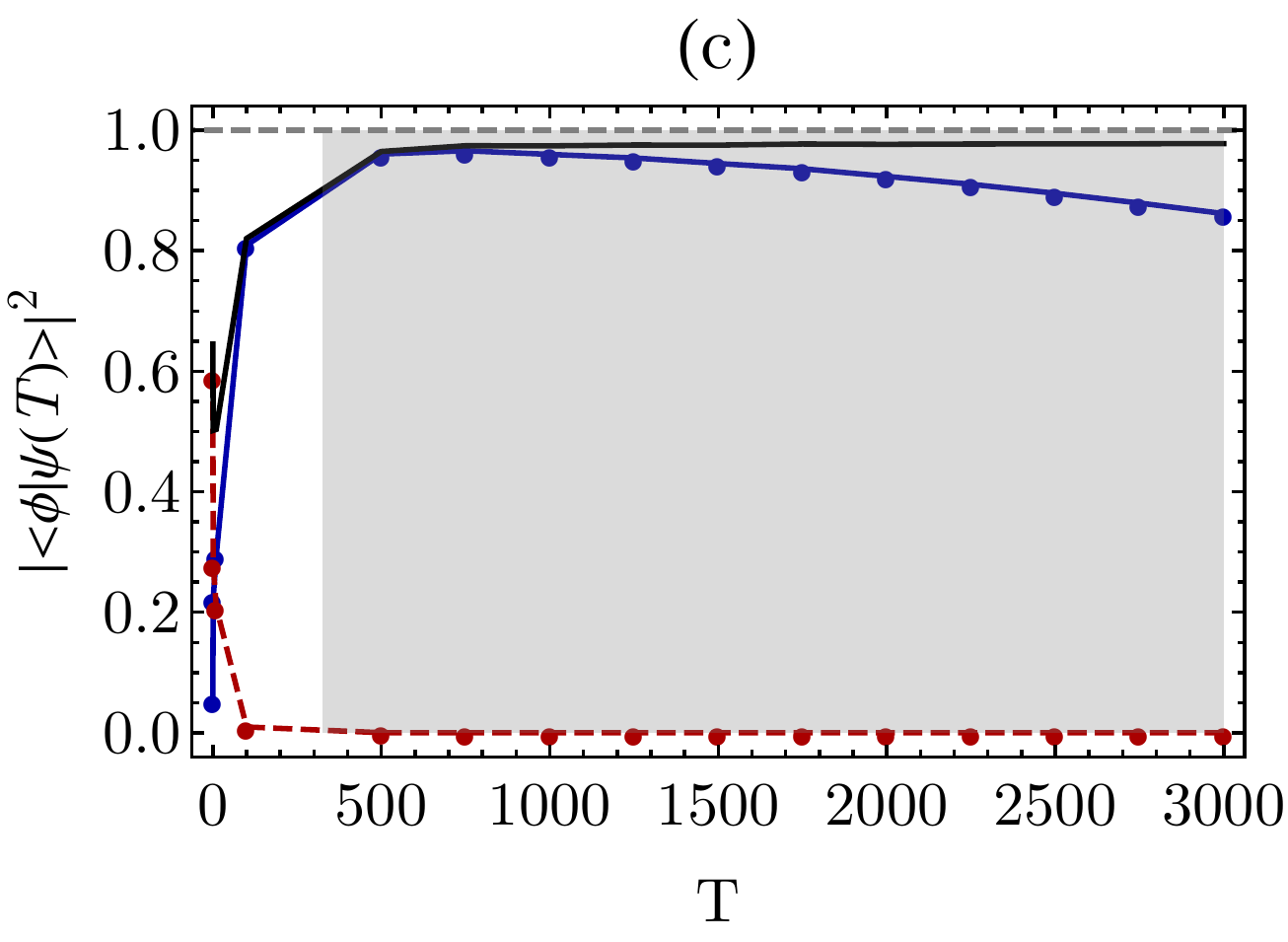}
\caption{(a) Gap to excitations as a function of half-angle $\theta$ for tuning forks with parameters $\eta=0.25$ (blue, solid), $\eta=0.5$ (yellow, dashed), $\eta=0.75$ (green, dotted), and $\eta=1$ (red, dot-dashed). Vertical grey dashed lines indicate the value of $\theta$ which maximises the gap for $\eta=1$ and $\eta=0.25$ ($\theta_c\approx20^\circ$ and $\theta_c\approx23^\circ$, respectively). (b) Majorana energy splitting at $s=0.125$ for tuning forks with parameters $\eta=0.25$ (blue, solid), $\eta=0.5$ (yellow, dashed), $\eta=0.75$ (green, dotted), $\eta=1$ (red, dot-dashed), shown on a logarithmic scale. (c) Final square overlaps for a tuning fork braiding simulation for $\eta=0.25$ and close to optimal half-angle $\theta=23^\circ$, to be compared with Fig.~\ref{fig:braiding_simulations}. Black solid line (without points) shows final square overlap with complete Majorana subspace, blue solid line (with points) shows the square overlap with the ideal final state, and red dashed line (with points) shows the overlap with the starting state. Dark grey shading indicates values of $T$ for which the braid was successful. Note that the dynamical oscillations here occur over much longer time scales than for the Y-junction geometry, and are barely visible in the figure. \label{fig:tuning_fork_simulations}}
\end{figure*}

The low success rate at small $T$ is in fact a reduction of the square overlap with the \emph{entire} MZM subspace, indicated by the black lines in Fig.~\ref{fig:braiding_simulations}. This small-$T$ feature shows that the final braided state lies outside of the initial low-energy subspace, and is therefore due to diabatic excitations to higher energy states that arise during the braid if it is performed too quickly. Indeed, the overlap with the whole MZM subspace is highest for $\theta=20^\circ$, which has the largest energy gap to excitations. Interestingly, the equivalent overlap for $\theta=25^\circ$ is higher than that for $\theta=15^\circ$, even though the excitation gap is smaller (see Fig.~\ref{fig:numerical_energies}(b)): this perhaps suggests that the braided state is more strongly excited into incipient Majorana modes than into generic bulk states. We also note that at very small $T$, the overlap with the MZM subspace initially \emph{decreases} for each value of $\theta$. This is due to the braid being so fast that most of the weight of the braided state $\ket{\psi(T)}$ remains in the initial state $\ket{1_-}$ (which is itself in the MZM subspace), as the system does not have enough time to adjust.

We now turn to the oscillatory behaviour of the overlaps with $\ket{1_-}$ and $\ket{1_+}$, shown in red and blue, respectively, in Fig.~\ref{fig:braiding_simulations} (and emphasised by the dark grey shading). These indicate oscillations of amplitude between the different states \emph{within} the (quasi)-zero-energy subspace (and includes smaller amplitude oscillations with states $\ket{2_+}$ and $\ket{2_-}$, not shown). These arise due to the fact that the four MZM states are not exactly degenerate at zero energy throughout the braid. Since the primary oscillation period decreases as $\theta$ increases, they must be associated with an energy splitting that increases as a function of $\theta$.

In Fig.~\ref{fig:numerical_energies}(c) we identified the largest such (constant) energy splitting, arising for $0\leq s\leq0.4$ in Fig.~\ref{fig:braid_bands}, which indeed increases exponentially with $\theta$. We can use this to obtain a heuristic explanation for the oscillatory behaviour. First, we recall that the braid we have simulated begins in state $\ket{1_-}$, which is an equal superposition of Majorana fermions $\gamma_1$ and $\gamma_2$. Halfway through the braid (i.e. at $s=1$ in Fig.~\ref{fig:braid_bands}), we have exchanged Majoranas $\gamma_2$ and $\gamma_3$, and so in the ideal case, the instantaneous state should be an equal superposition of the original $\gamma_1$ and a MZM located where $\gamma_3$ was initially. We can write this instantaneous state in terms of the initial states as
\bequ
\ket{\psi(T/2)}=\frac{1}{2}\bigg[\left(\ket{1_-}+\ket{1_+}\right)+e^{i\phi}\left(\ket{2_-}+\ket{2_+}\right)\bigg],
\eequ
where $\phi$ is the (unspecified) phase difference between the two pairs of terms. The combination of states $\ket{1_-}$ and $\ket{1_+}$ corresponds to $\gamma_1$, while the combination of states $\ket{2_-}$ and $\ket{2_+}$ corresponds to $\gamma_3$. 

For the next section of the drive (corresponding approximately to $1\leq s\leq1.4$), the instantaneous state is not an eigenstate, and so different terms in the wavefunction expansion will pick up different dynamical phases. Assuming the instantaneous energy levels are constant during this time (i.e. ignoring the oscillatory pieces in Fig.~\ref{fig:braid_bands}), we can write the evolved state as
\beq
\ket{\psi(T/2+t)}&=&\frac{1}{2}e^{iE_1t}\bigg[\left(\ket{1_-}+e^{-2iE_1t}\ket{1_+}\right)\\
&&+e^{i\phi+i(E_2-E_1)t}\left(\ket{2_-}+e^{-2iE_2t}\ket{2_+}\right)\bigg],\nonumber
\eeq
where have defined $E_1=E_{1_+}=-E_{1_-}$ and $E_2=E_{2_+}=-E_{2_-}$ as the energies of the corresponding eigenstates. From Fig.~\ref{fig:braid_bands}, the energy $E_2$ is the dominant energy scale, and will cause a shift in the relative phase of $\ket{2_-}$ and $\ket{2_+}$. This will cause oscillations between Majorana modes $\gamma_3$ and $\gamma_4$, which will in turn alter the final state of the braid. 

This picture suggests that the oscillations will have a dominant characteristic time period of approximately $T'=\pi/E_2$, which is comparable with the numerical oscillations observed in the braiding overlaps. Specifically, extracting the MZM splitting from the data in Fig.~\ref{fig:numerical_energies}, we obtain predicted time periods of $T_1\approx 4800, 1800, 650$ for $\theta=15^\circ, 20^\circ, 25^\circ$, respectively. Extracting the time periods from Fig.~\ref{fig:braiding_simulations} directly instead gives $T_1\approx 3000, 1250, 440$. The differences likely arise from the oscillatory regions of the MZM energy splitting (where the energy levels are changing with time), and from other energy splittings that arise during the braid. In addition to this behaviour with period $T_1$, other oscillations with approximate time period $T_2=2\pi/(E_2-E_1)\approx 2T_1$ are also visible in the overlap with states $\ket{2_-}$ and $\ket{2_+}$ (not shown).

Overall, these numerical simulations suggest that a topological braiding operation is feasible in a realistic nanowire device, but that dynamical phases may need to be accounted for carefully. Notably, the final braided state remains within the low-energy subspace for a total braid duration of approximately $T\gtrsim1000$ in our units. This corresponds to $T\gtrsim4$~ns, which compares favourably to measured quasiparticle poisoning times which are typically on the order of microseconds \cite{HigginbothamParity2015,AlbrechtTransport2017}.

The oscillatory behaviour of the final state overlaps demonstrates the difficulties that may arise from dynamical phases. To use the simulated devices directly, it would likely be necessary to sweep the braiding time $T$ to find a maximum braiding success probability. This may require repeated measurements to determine the complete profile of the overlap curves, and may be complemented by performing repeated braiding operations (as suggested in Ref.~\onlinecite{ClarkeProbability2017}). More generally, these dynamical phase effects can be minimised by reducing the energy splitting between Majorana modes. In particular, the most significant splitting arises from the two Majorana modes localised in the angled prong of the Y-junction at the beginning of the braid. This can be reduced by increasing the length of the device legs (so that the Majorana modes are further apart) or by increasing the effective superconducting gap (to reduce the extent of the Majorana modes). However, it may be more experimentally favourable to make just the lower-right leg of the device longer (possibly at the expense of the others), so that the most significant energy splitting is reduced. It may also be advantageous to use a narrower prong angle $\theta$, which reduces the Majorana energy splitting (albeit at the expense of a smaller gap to excitations). Finally, the critical angle $\theta_c$ is known to increase as the ratio $\Delta_0/V_Z$ approaches one \cite{RexTilting2014} (at the expense of a smaller band gap at $k=0$); there may therefore be some advantages to reducing the external magnetic field.

\section{Tuning Forks and Other Parameter Regimes\label{sec:tuning_forks}}

In the discussions above, we considered a Y-junction device with experimentally motivated `best case' parameters introduced in Sec.~\ref{sec:system_setup}. In this section, we discuss the benefits that may arise from instead using a tuning fork geometry [as in Fig.~\ref{fig:ytuningfork}(b)]. We also consider how robust our conclusions are to changes in the underlying parameters.

In Fig.~\ref{fig:tuning_fork_simulations}(a), we show the minimum gap to excited energy states (during the braid) as a function of half-angle $\theta$, for a variety of choices for $\eta$, the geometric parameter that determines when the tuning fork straightens out (see Fig.~\ref{fig:ytuningfork}(b)). We see that as $\eta$ decreases, and more of the device becomes aligned with the magnetic field, the larger the maximum gap becomes (and the larger the value of $\theta$ is for which this occurs). This arises because the tuning fork geometry has higher energy bulk states as compared to the Y-junction: more of the device is aligned with the magnetic field, and so the band tilting effect shown in Fig.~\ref{fig:numerical_energies}(a) is less significant. In this way, the maximum excitation gap arises for larger values of $\theta$, when the effects of the bulk gap closure and incipient Majorana modes are equal. For $\eta=0.25$, the maximum gap arises at $\theta\approx 23^\circ$.

The tuning fork geometry also leads to smaller energy splitting between the Majorana modes, as shown (for $s=0.125$ during the braid) in Fig.~\ref{fig:tuning_fork_simulations}(b). This is because the splitting is exponentially small in the Majorana localisation length, which in turn is inversely related to the bulk energy gap. For $\eta=0.25$, the energy splitting is almost two orders of magnitude smaller than for the Y-junction ($\eta=1$).

These two features suggest that tuning fork devices are more promising geometries for realising and braiding Majorana modes than Y-junctions. The larger bulk gap means that the braid should not need to be performed as slowly to be in the adiabatic regime, and the reduced Majorana energy splitting should reduce the unhelpful effects of dynamical phases. Indeed, in Fig.~\ref{fig:tuning_fork_simulations}(c) we plot the final braiding overlaps for the tuning fork geometry and see that both of these advantages are apparent: The overlap with the Majorana subspace is higher for shorter braiding times $T$ and the dominant period of oscillation is significantly longer, and is barely visible in the figure. Overall, the final braided state is close to the ideal state for almost all of the simulated braiding times. This suggests that the tuning fork geometry has significant advantages over the Y-junction geometry.

Unfortunately, we cannot improve these energy scales indefinitely by taking $\eta\to0$. Not only would this pathological limit be impossible to engineer, but it would also lead exactly to a $\pi$-junction forming at the node, which would interfere with the braiding process. The optimum value of $\eta$ is likely to depend on the spatial extent of the Majorana modes: when two Majorana modes meet at the junction, they should have most of their weight on an angled part of the device, to avoid forming a $\pi$-junction.

In passing, we note that the tuning fork geometry is likely to have an additional advantage over the Y-junction geometry in terms of orbital effects, which lie beyond the scope of the one-dimensional model. Specifically, Peierls phases are expected to be gained by electrons hopping perpendicularly to the external magnetic field, which is most significant in the angled legs of the device. According to Ref.~\onlinecite{NijholtOrbital2016}, this is expected to reduce the gap to excitations in a similar manner to the interplay between the Zeeman term and spin-orbit coupling. This effect should be smaller for the tuning fork geometry, which has more of the device aligned with the external field. A detailed study of the orbital effect of the magnetic field in three-legged devices (using 3D simulations) remains an interesting avenue for future research.

\begin{figure}
\includegraphics[clip=true, trim = 0 25 0 0, scale=0.5]{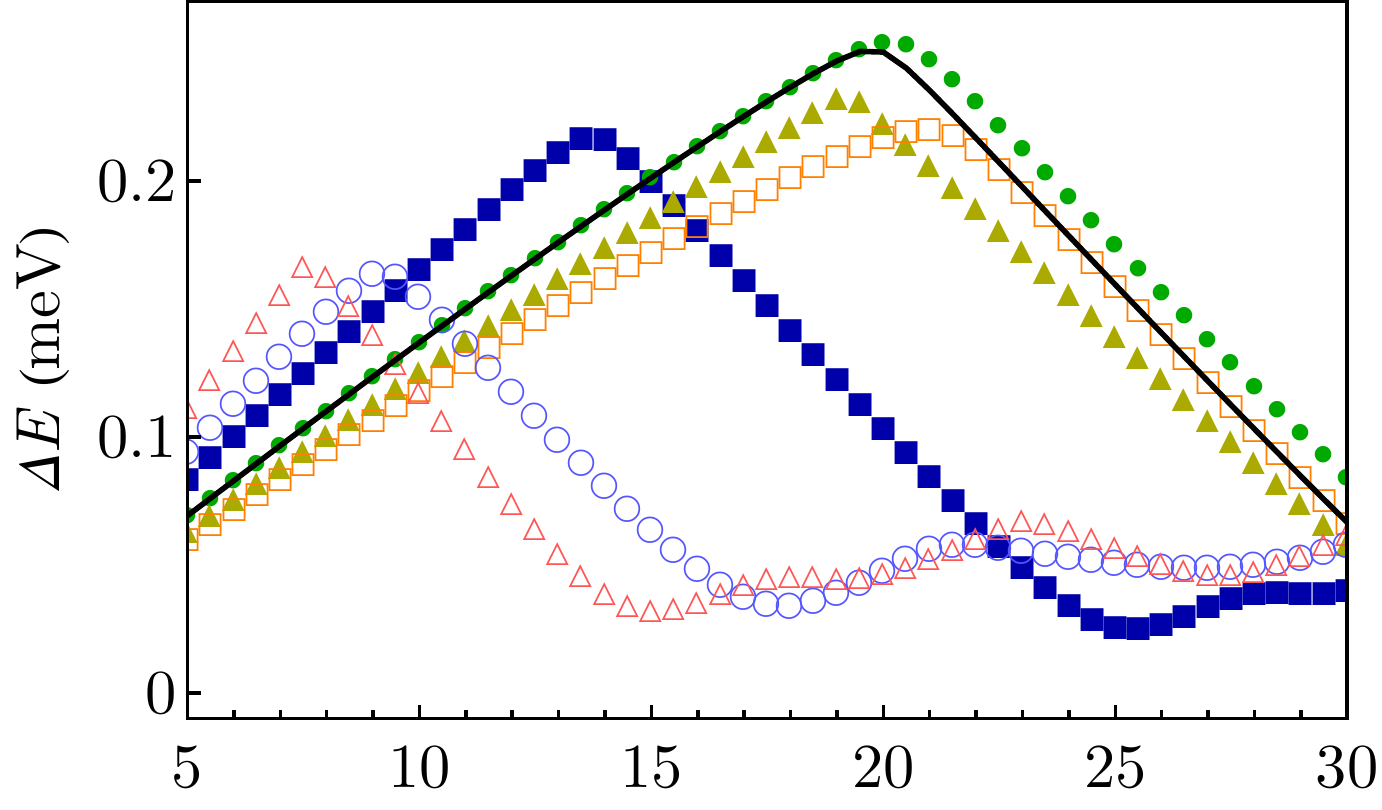}
\includegraphics[clip=true, trim = 0 0 0 0, scale=0.5]{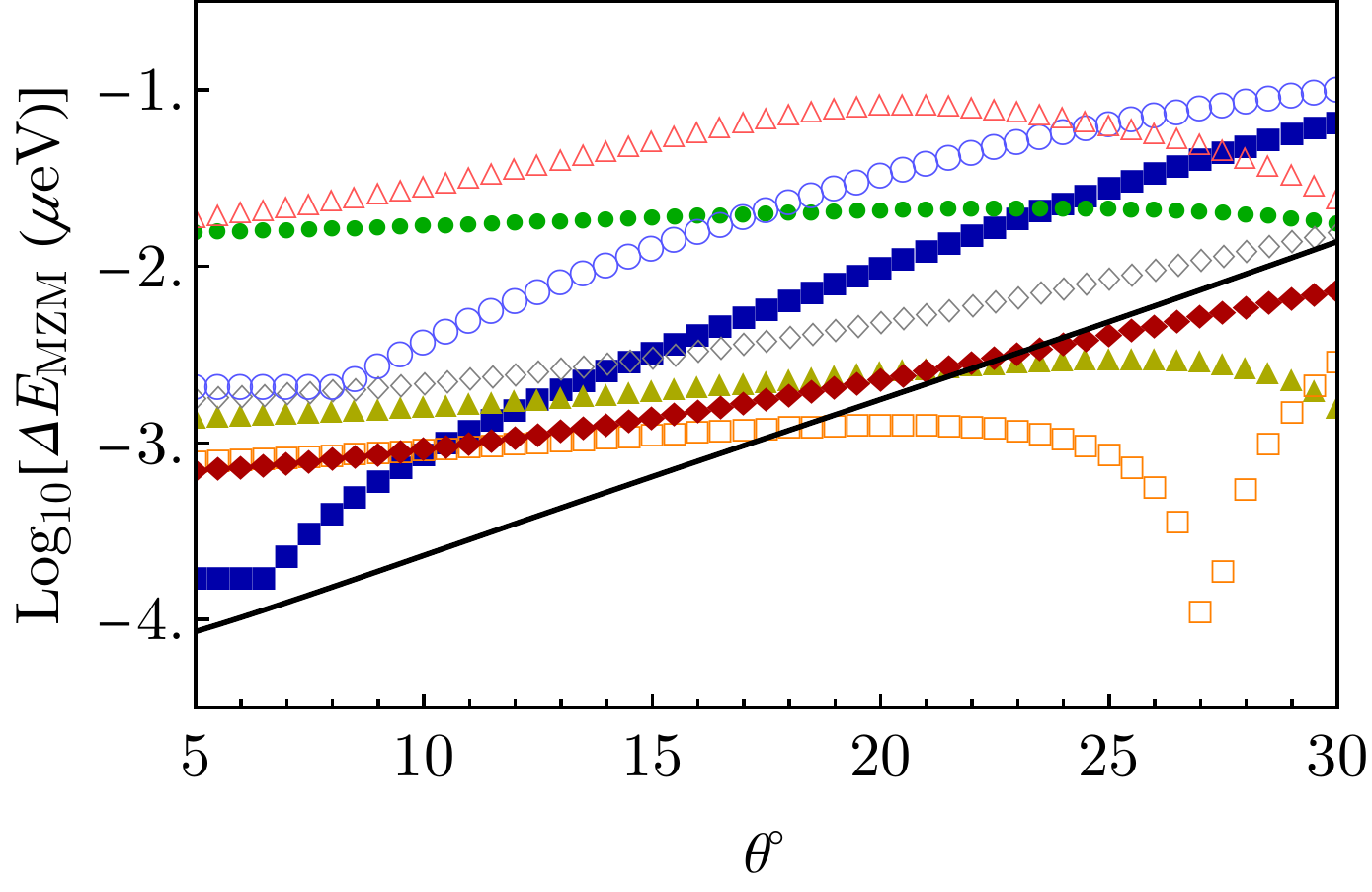}
\caption{Excitation gap (top) and Majorana splitting energy (bottom) for different model parameter choices as a function of $\theta$ for a Y-junction device. Black solid line shows `best case' parameters described in Sec.~\ref{sec:system_setup}. The other symbols each have a parameter alteration: $\Delta_0=0.6~{\rm meV}$ (dark blue filled squares); $\Delta_0=0.4~{\rm meV}$ (light blue open circles); $L=1.5~\mu{\rm m}$ (green filled circles); $\alpha_R=0.35~{\rm meV}$ (yellow filled triangles); $\mu_{\rm triv}=3~{\rm meV}$ (orange open squares); $\beta=2.5~(\mu{\rm m})^{-1}$ (dark red filled diamonds); $\beta=10~(\mu{\rm m})^{-1}$ (grey open diamonds); doubling of the magnetic field strength so that $V_Z=3$~meV and $\Delta_0\to0.68$~meV (pink open triangles). Note that the dark red and grey symbols overlap with the best case parameter line in the upper panel, and are not shown. See main text for details. \label{fig:params}}
\end{figure}

In order to study the robustness of our conclusions, we now relax our best-case parameter choices and observe how the band gap and Majorana energy splitting vary. Specifically, we alter the induced pairing gap, the length of the device legs, the spin-orbit coupling strength, the external magnetic field, and aspects of the braiding protocol. The results are plotted in Fig.~\ref{fig:params}. 

We see that in general, modifying these parameters away from their `best case' values decreases the size of the bulk gap (upper panel of Fig.~\ref{fig:params}) and increases the value of the Majorana energy splitting (lower panel of Fig.~\ref{fig:params}). The most significant parameters affecting the Majorana splitting are the induced superconducting pairing gap $\Delta_0$, the length of the device legs $L$, and the external magnetic field strength $B$. Changes in $\Delta_0$ or $L$ of about 25\% increase the Majorana splitting energy by about one order of magnitude. Modifying the other parameters (notably changing the steepness and strength of the chemical potential profile) has a comparatively smaller effect on the splitting.

A change in the magnetic field strength $B$ affects the Hamiltonian in two important ways: first, it directly changes the Zeeman energy through the relation $V_Z=\frac{1}{2}g_{\rm eff}\mu_BB$, and secondly, it indirectly reduces the superconducting pairing energy as the external magnetic field approaches the critical field of the superconductor, $B_c$. We incorporate this suppression through the relation
\beq
\Delta_0(B)&=&\Delta_0\sqrt{1-\left(B/B_c\right)^2},
\eeq
which reproduces the pairing collapse observed in experiments well \cite{ZhangQuantized2018}. For the nitride-based superconductors we have assumed in this text, the critical field strength is about $B_c=5$~T \cite{PushpPrivate2018}, which we have used in Fig.~\ref{fig:params}. Specifically, for the pink curve in Fig.~\ref{fig:params}, we have increased the field from $B\approx1.3$~T to $B\approx2.6$~T. This changes the Zeeman energy from $V_Z=1.5$~meV to $V_Z=3.0$~meV and reduces the pairing strength from $\Delta_0=0.8$ to $\Delta_0=0.68$. As a result, the system has a much larger Majorana energy splitting and a substantially reduced bulk gap. From these energetic considerations, it seems sensible to choose an external magnetic field which is large enough for the system to be in the topological regime, but not so large that the pairing gap starts to collapse significantly. 

We note from Fig.~\ref{fig:params} that the bulk gap is more robust to changes in the underlying parameters than the Majorana energy splitting, although the precise value of the optimum half-angle $\theta_c$ is slightly different in each case. The most dramatic changes arise when the superconducting pairing gap $\Delta_0$ is reduced or the magnetic field strength $B$ is increased, with a smaller bulk energy gap having a significant effect on the overall gap profile as a function of $\theta$. This would be particularly important for devices with aluminium superconductors, whose pairing gap is usually taken to be about $0.2~{\rm meV}$, and whose critical field $B_c$ is close to $1.1$~T \cite{LiuAndreev2017,ZhangQuantized2018}, both significantly lower than for nitride superconductors.

Following the plots in Fig.~\ref{fig:params}, we would expect suitable aluminium-based devices to require a much narrower half-angle (in the range $5^\circ$--$10^\circ$). Perhaps surprisingly, however, the magnitude of the maximum gap does not decrease linearly with $\Delta_0$, and so aluminium-based devices should still be suitable for braiding operations. In addition, the smaller required values of $\theta$ should also suppress the Majorana splitting, to partially compensate for the increase in localisation length due to the smaller bulk gap. These features could likely also be further improved by using a tuning fork geometry.

Overall, these results suggest that our conclusions on the feasibility of braiding are reasonably robust to changes in the parameters of the underlying model. From an experimental perspective, it would seem most important to achieve a large induced superconducting gap and long device legs (although this latter condition is likely to amplify environmental effects). Variations in the model parameters affect the Majorana energy splitting more strongly than the overall gap to excitations, suggesting that the effects of dynamical phases will differ from device to device and may need to be characterised carefully in each case. However, these issues are likely to be improved if a tuning fork geometry is used instead of a pure Y-junction, following the discussion above.

\section{Conclusion\label{sec:conclusion}}
In this work, we have studied the feasibility of performing braiding operations in nanowire devices in the shape of Y-junctions and tuning forks. As a stepping stone towards understanding a real braiding experiment, we have used a continuum model to describe these devices, with parameters extracted from existing transport experiments. This model has previously been shown to give good agreement with experimental measurements.

\begin{figure*}[t]
\includegraphics[scale=0.32, clip=true, trim = 0 -4 0 0]{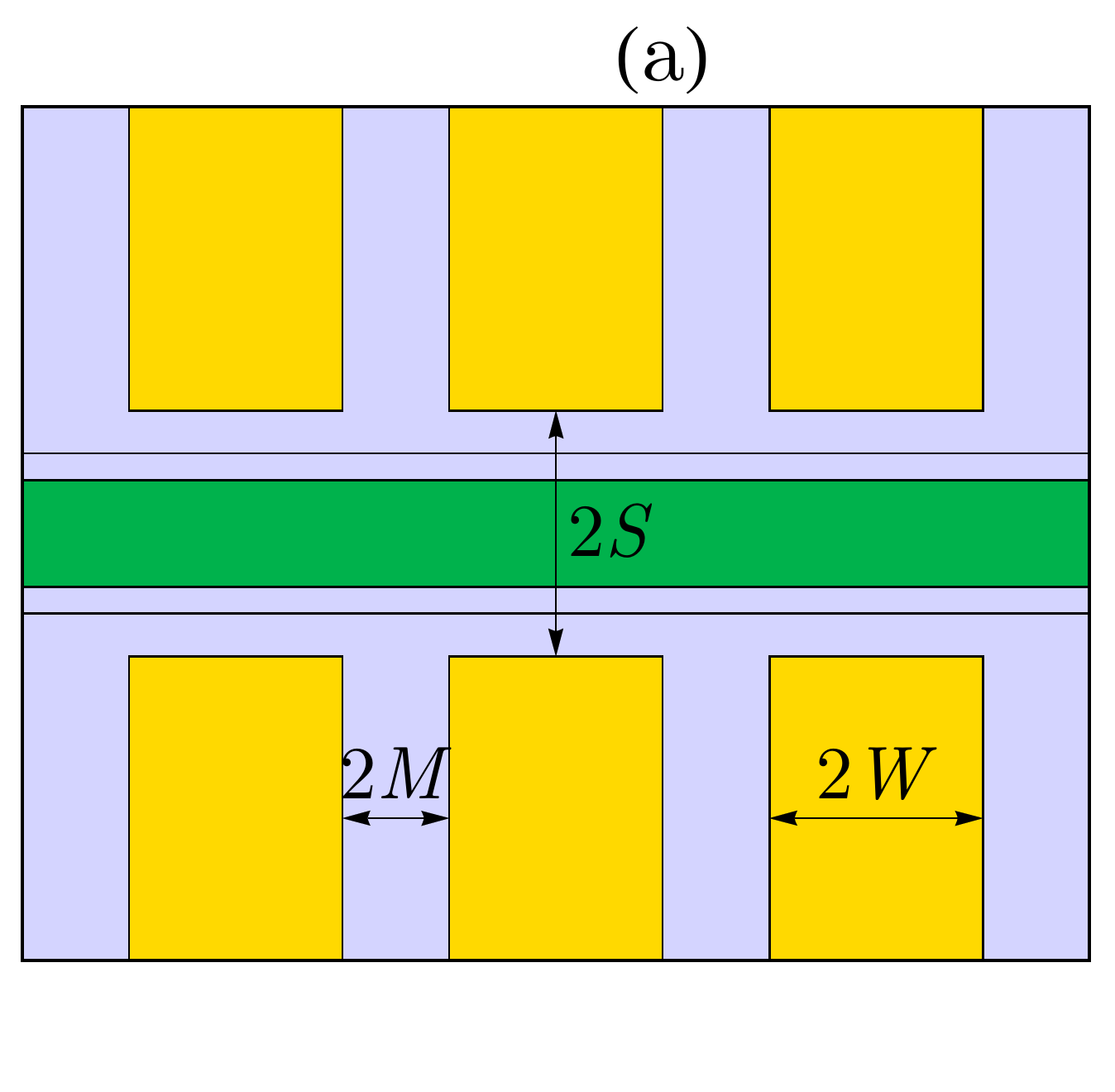}
\includegraphics[scale=0.32, clip=true, trim = 0 -10 0 0]{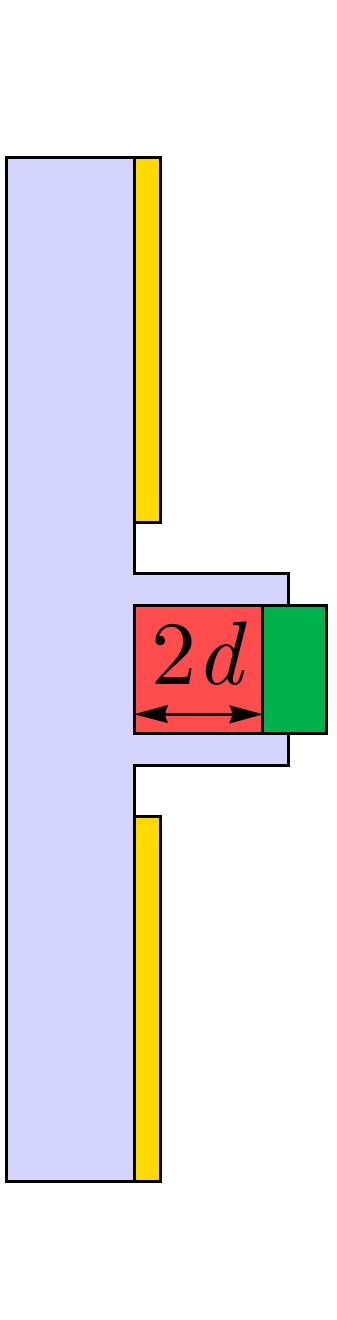}
\hspace{3mm}
\includegraphics[scale=0.32]{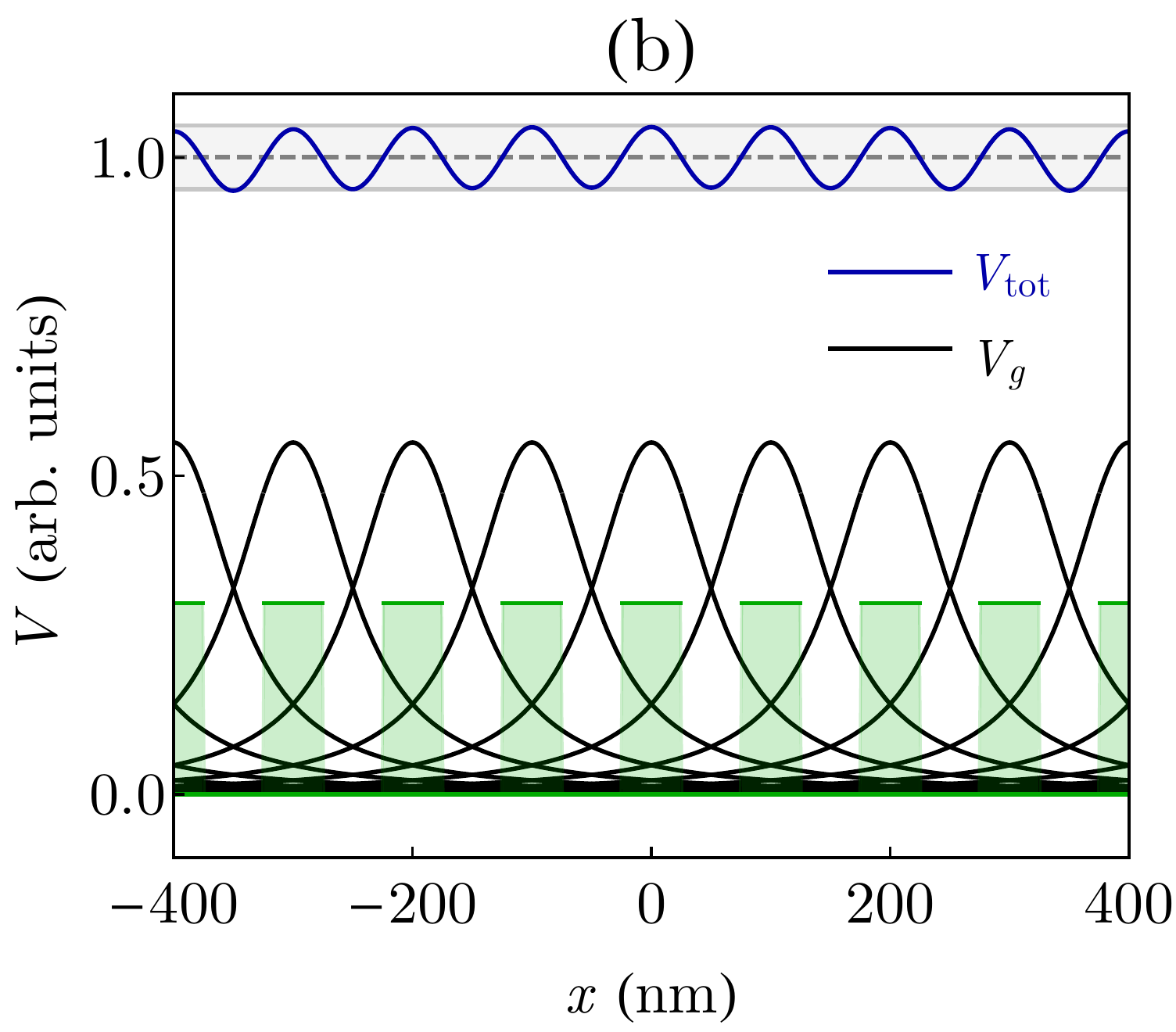}
\hspace{3mm}
\includegraphics[scale=0.32]{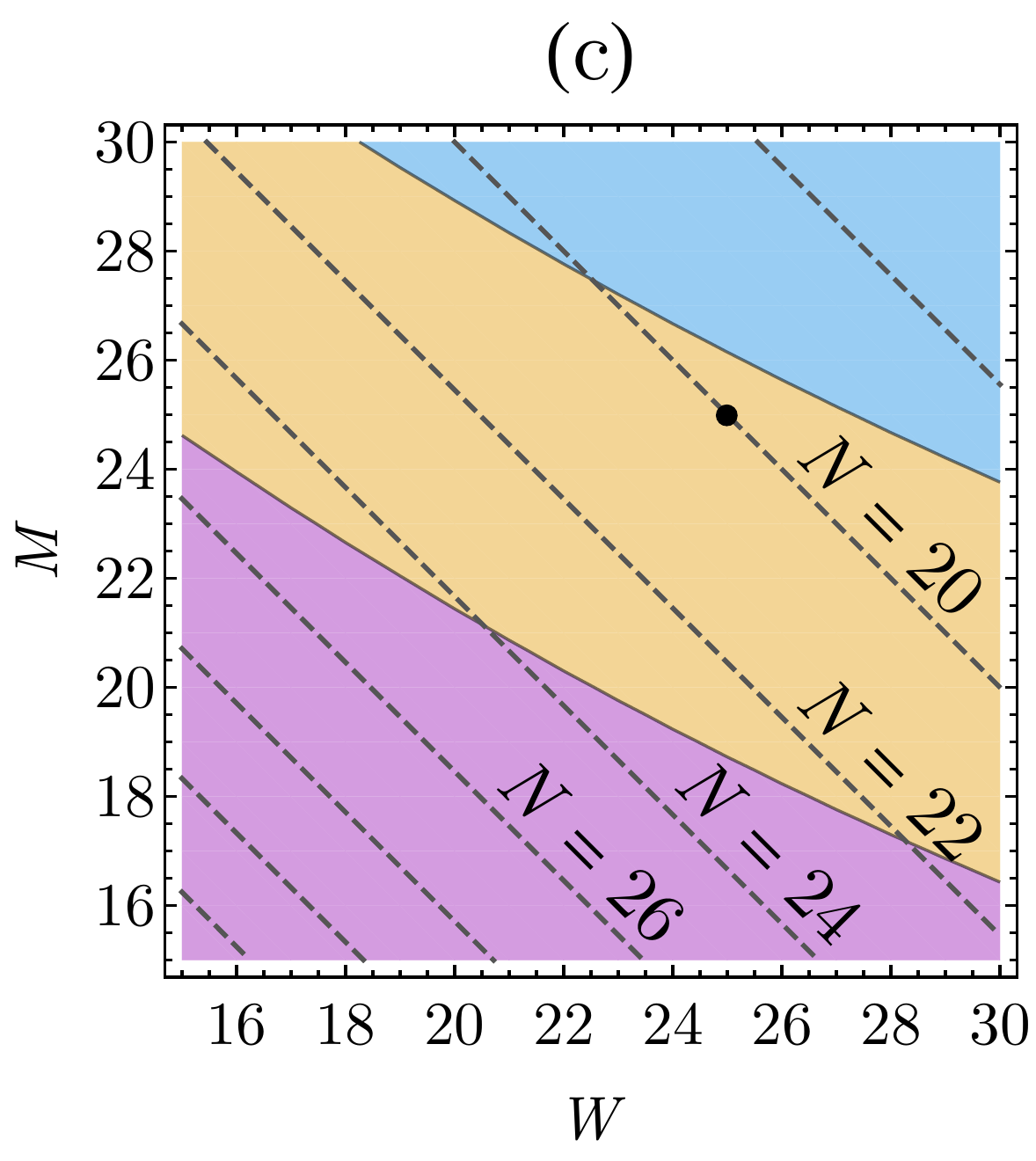}
\includegraphics[scale=0.27, clip=true, trim = 20 -50 80 0]{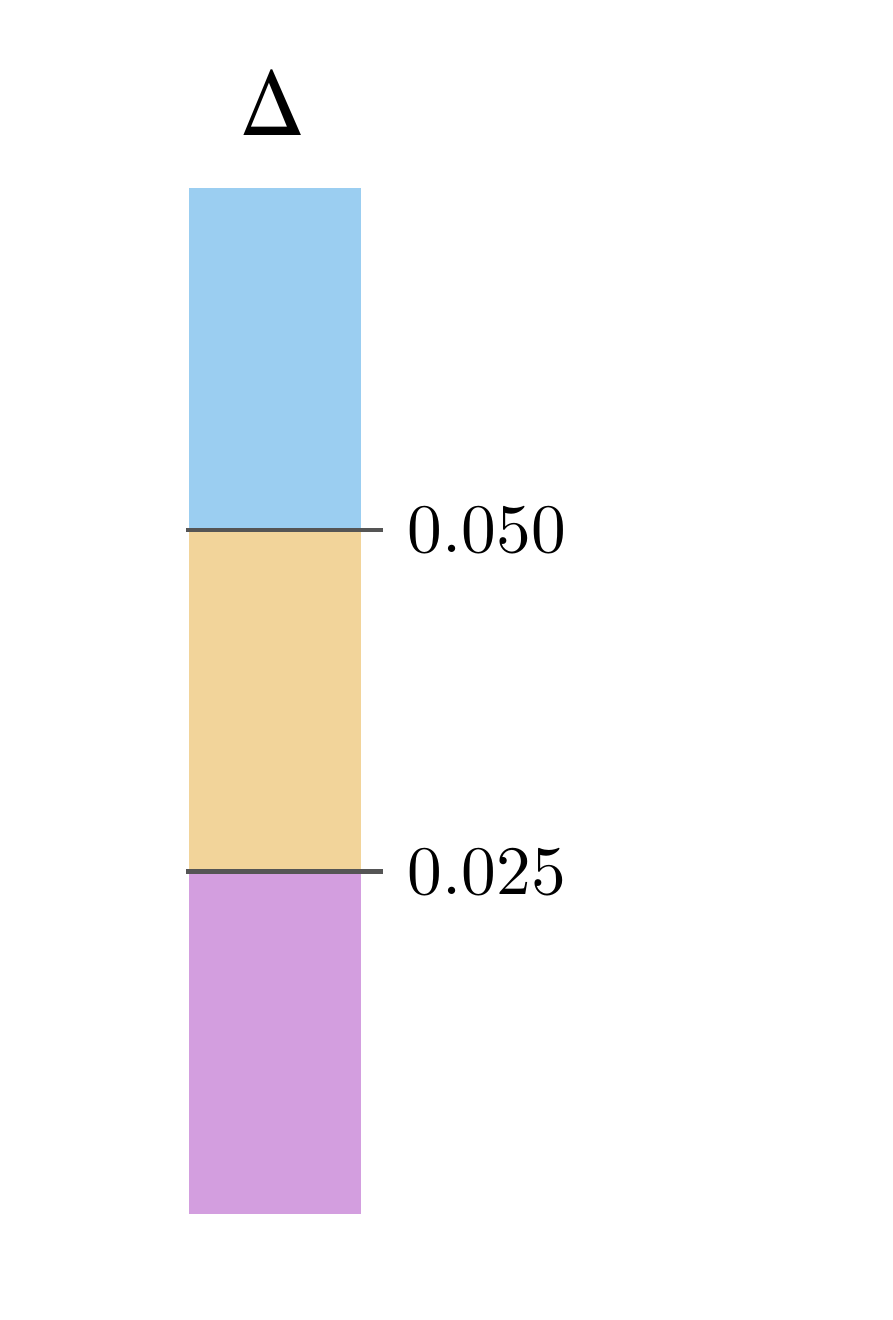}
\caption{(a) A section of a nanowire device with the gate arrangement and geometry considered in Appendix~\ref{app:gates}, showing both a plan (left) and a cross section (right). The nanowire is shown in red with a green superconducting layer on top (running through the centre of the device), while the gates themselves are yellow (above and below the nanowire). We take $d=20~{\rm nm}$ and $S=40~{\rm nm}$, and allow $W$ and $M$ to vary. (b) A series of gates with $W=25~{\rm nm}$ and $M=25~{\rm nm}$ [green (gray) rectangles, below], showing the potential from each gate (black solid lines, below) and the total potential for an infinite series of gates (dark blue solid line, above). The grey band shows the vertical interval $[0.95,1.05]$, which is roughly the same scale as the oscillations of the total potential. (c) A region of the $W$-$M$ plane indicating the value of the fluctuation ratio $\Delta$, with (solid) contours at $\Delta=0.05$ and $\Delta=0.025$ (where $\Delta$ increases towards the top-right corner of the plot). Dashed overlaid contours show the number of gates $N$ that would be found on a single $2~{\rm \mu m}$ leg of a nanowire device considered in the main text (only even $N$ shown). The black dot shows the point $W=25~{\rm nm}$, $M=25~{\rm nm}$, which are the parameters used in (b) and elsewhere in the text.\label{fig:gates}}
\end{figure*}

We began by computing the energetic properties of a Y-junction as a putative braid is performed, noting that the gap to excitations, which affects the speed at which a braid can be performed, depends on two factors. First, the bulk gap may collapse due to the misalignment of the external magnetic field and the effective spin-orbit coupling field. This effect is most prominent when the Y-junction makes a wide angle to the external field. However, if the angle between the legs is too narrow, then incipient Majorana modes may form at the junction, which also reduces the gap to excitations. The optimum angle which maximises this gap depends on the device parameters, but for our choice of values was close to $\theta_c\approx20^\circ$.

A realistic device also exhibits energy splitting between low-lying Majorana modes. This is determined by the localisation length of the Majorana modes, which in turn is inversely related to the bulk gap. The energy splitting was found to increase exponentially with the Y-junction angle, in approximate correspondence with the closure of the bulk gap.

Both of these static considerations have important implications for the feasibility of a real braiding operation. Notably, a smaller gap to excitations means that the braid must be performed more slowly, while nonzero Majorana splitting generates dynamical phases that produce oscillations in the final braided state. We studied these effects in detail by simulating a braiding operation for a Y-junction with three different values of $\theta$ and calculating the overlap of the final state with the theoretical prediction. Overall, our results suggest that such devices are capable of performing nonabelian braiding operations, but that dynamical phase effects may need to be carefully taken into account. In order to be successful, the braiding operation should take place over a time scale of at least 4~ns, which compares favourably to existing estimates of quasiparticle poisoning times \cite{HigginbothamParity2015,AlbrechtTransport2017}.

In Sec.~\ref{sec:tuning_forks}, we showed that a tuning fork geometry has significant advantages over the Y-junction geometry. By increasing the proportion of the nanowire that is aligned with the external magnetic field, the gap to excitations can be increased and the Majorana energy splitting reduced. In turn, this reduces the prominence of dynamical phase oscillations, and means that dynamical excitations to higher energy states are suppressed, increasing the speed with which the braid can be successfully performed. Tuning forks are also expected to be more robust to orbital field effects. Beyond this, we studied the robustness of our simulations to small variations in the parameters of our underlying model, finding that the induced pairing gap, and to a lesser extent the length of the device, are the most important factors in determining the success of a braid.

Our numerical results provide qualitative and quantitative statements about the feasibility of braiding in realistic nanowire devices, which we believe are overall encouraging. In particular, our simulations suggest that such a braid lies within experimental capabilities, although many steps undoubtedly remain before this becomes a reality. To further bridge this divide between theory and experiment, a number of extensions to this numerical work could be made. First, while we have based our calculations on a model which has previously shown good agreement with experiments, it remains a one-dimensional model that is incapable of capturing certain effects (notably the effect of subbands and the orbital effect of the field). It would be interesting to extend this type of simulation to a full three-dimensional model which includes these extra features, and perhaps more realistic gating setups. In addition, we introduced superconducting pairing to the device through a bare pairing term in the Hamiltonian. There are now a number of works which have introduced methods to carefully model the superconductor-nanowire interface (see, for example, Ref.~\onlinecite{Winklerunified2018}), which could be incorporated into a simulation of this kind. Finally, there are many different mechanisms which affect the lifetime of a Majorana qubit (see, for example, Refs.~\onlinecite{RainisMajorana2012a,AseevLifetime2018a}). A direct calculation of such lifetimes for a specific Y-junction geometry would allow the feasibility of a braiding operation to be more sharply evaluated. Ultimately, of course, the feasibility of braiding as a tool for quantum computation can only be confirmed by a successful experiment.

\begin{acknowledgments}
We gratefully acknowledge discussions and ongoing collaborations with B.~Madon, M.~A.~Mueed, F.~Nichele, H.~Riel and M.~Ritter. We also thank S.~Rex and J.~Perk for a number of very useful comments on the manuscript. This work was supported by DARPA Topological Excitations in Electronics (TEE) project No.~140D6318C0028. The authors also acknowledge computational resources supported by the NSF under CAREER DMR-1455368.
\end{acknowledgments}

\appendix
\section{Side Gate Arrangement\label{app:gates}}
In this appendix, we show that a chemical potential ramp profile similar to the one used in our simulations could be achieved using a realistic arrangement of side gates. In particular, a rough calculation suggests that a gate density of approximately one gate every 100~nm should be sufficient to realise a ramp profile to within fluctuations of approximately 5\%. For the devices we consider in the main text, this corresponds to about twenty gates per leg. 

To simplify the calculation, we assume that the gates are all identical rectangular plates with width $2W$, and are arranged symmetrically about the nanowire device as shown in Fig.~\ref{fig:gates}a. Each pair of gates has a separation distance of $2S$, while neighbouring pairs are separated by a distance $2M$. Ignoring any screening effects, a pair of gates centred at $(x=x_j, y=0, z=0)$ and held at a potential $U_j$ generates an electrostatic potential energy at $(x, y, d)$ given by \cite{KjaergaardQuantized2016,DaviesModeling1995}
\begin{widetext}
\beq
V_{\rm g}(x,y,d;x_j,U_j)&=&\frac{(-eU_j)}{\pi}\left[\mathrm{arctan}\left(\frac{W+x-x_j}{d}\right)+\mathrm{arctan}\left(\frac{W-x+x_j}{d}\right)\right]-g(S+y,W+x-x_j)\nonumber\\
&&-g(S+y,W-x+x_j)-g(S-y,W+x-x_j)-g(S-y,W-x+x_j),\label{eq:v_gate}
\eeq
\end{widetext}
with
\beq
g(u,v)&=&\frac{1}{2\pi}\mathrm{arctan}\left[\frac{uv}{dR(u,v)}\right]
\eeq
and
\beq
R(u,v)&=&\sqrt{u^2+v^2+d^2}.
\eeq

We will calculate the total potential energy at the centre of the nanowire device, ignoring screening effects, by summing the potentials from a series of gate pairs held at different potentials $U_j$ and centred at different locations $x_j$. We will assume that the centre of the nanowire is at a height of $d=20~{\rm nm}$ (as is the case for nanowires grown using TASE), and that each pair of gates has a separation of $2S=80~{\rm nm}$, which is two nanowire widths. We will allow $W$ and $M$ to be varied. Since the overall energy scale can be altered using the finger gates and the back gate, we will try to find a gate arrangement that reproduces a ramp function between the arbitrary units of zero and one. To simplify the calculation we will use a linear ramp function (rather than the sine-squared function used in Eq.~\ref{eq:ramp_function}), but the results should be similar for any ramp function that varies over the same length scale.

We first consider oscillations in the electrostatic potential away from the ramp itself, in the region where the total potential should be constant (and, in our units, equal to one) [see Fig.~\ref{fig:gates}b for an illustration of these oscillations]. In this region, all gates will be held at the same potential $U_j=U$, and the total potential in the centre of the nanowire can be found by summing a series of potentials of the form in Eq.~\eqref{eq:v_gate}, each with a different offset $x_j=2j(W+M)$. The potential will vary as a function of $x$ along the nanowire device, taking its largest value at $x=x_j$ (aligned with the centre of a pair of gates) and a minimum value at $x=x_j+W+M$ (aligned with the midpoint between two neighbouring pairs of gates). Approximating the number of gates as infinite, we can write
\beq
V_{\rm max}&=&\sum_{j=-\infty}^{\infty}V_{\rm g}\left(x=0;x_j\right)\\
V_{\rm min}&=&\sum_{j=-\infty}^{\infty}V_{\rm g}\left(x=j(M+W);x_j\right),
\eeq
where we have suppressed the variables $y=0$, $d=20~{\rm nm}$ and $U_j=U$ which are the same for all gates, and where pairs of gates are centred at $x_j=2j(W+M)$.

\begin{figure}[t]
\includegraphics[scale=0.45, clip=true, trim = 0 28 0 0]{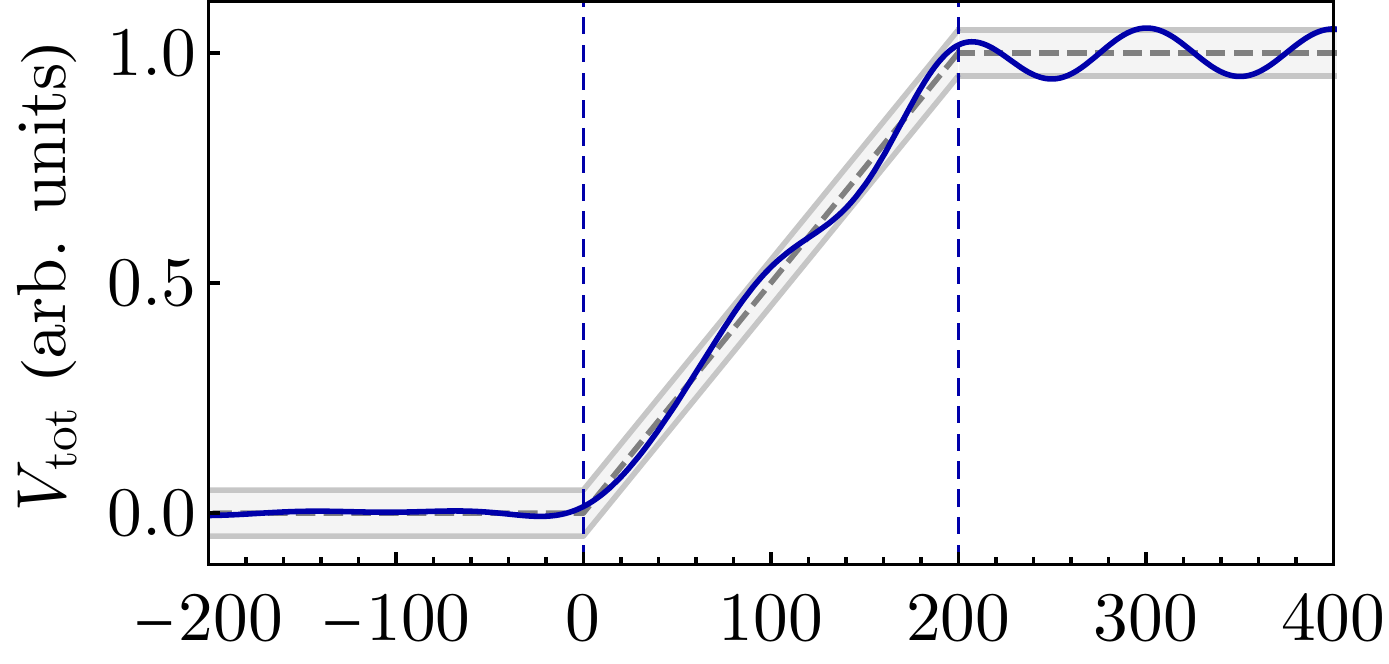}
\includegraphics[scale=0.465, clip=true, trim = 0 28 -1 0]{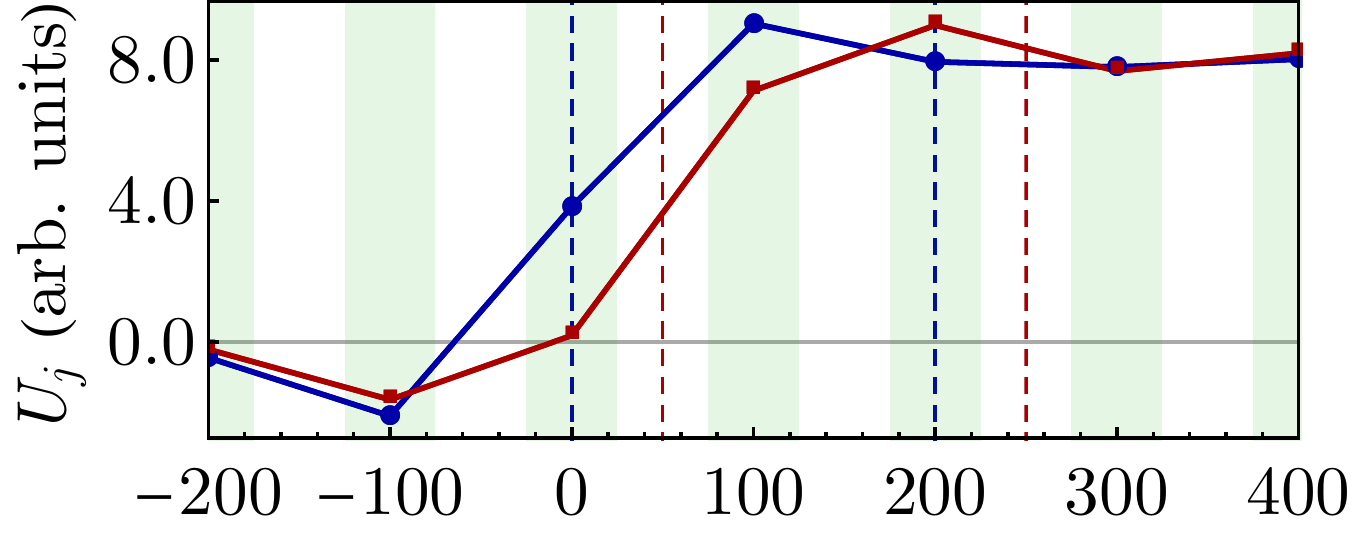}
\includegraphics[scale=0.451, clip=true, trim = 0 0 0 0]{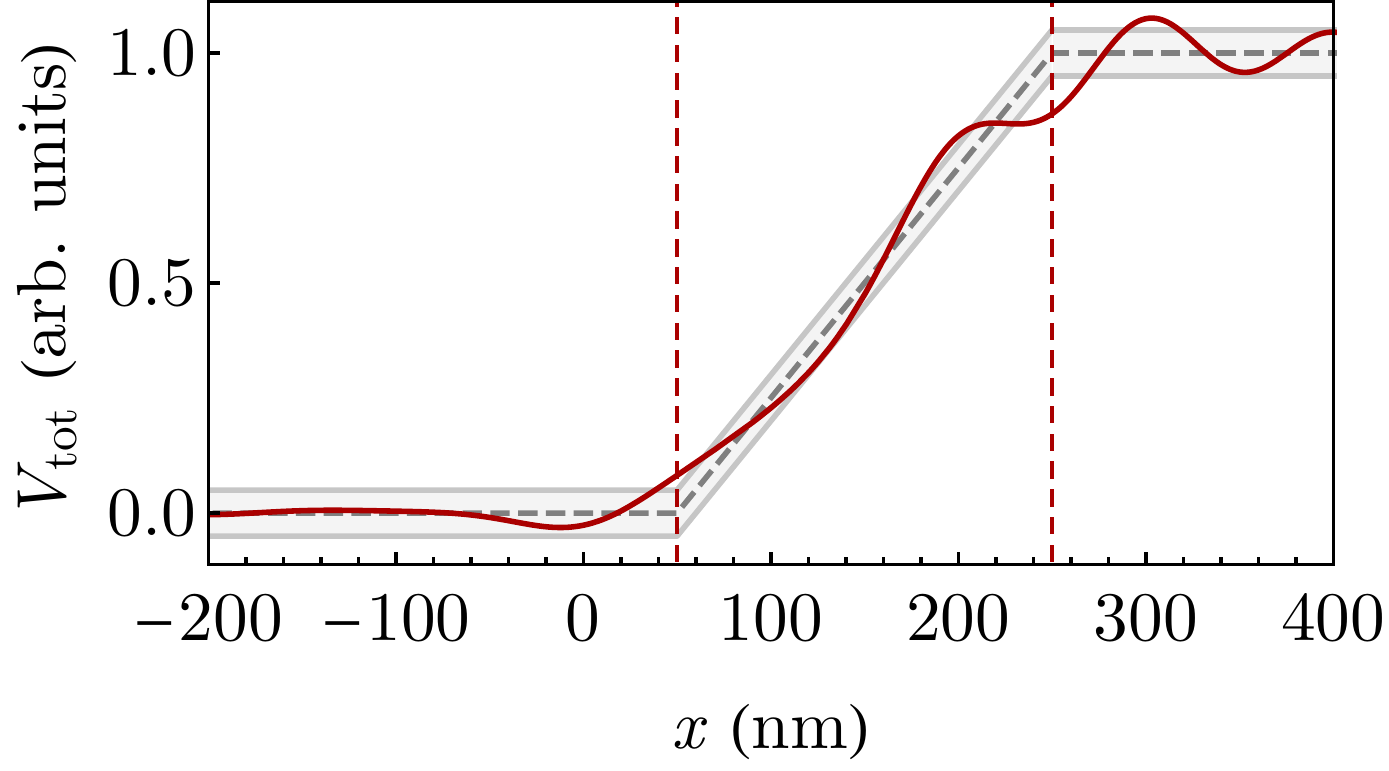}
\caption{Top: ideal ramp profile (grey dashed line) and optimised ramp profile generated by eleven side gates (dark blue solid line), normalised so that the ideal ramp peaks at one. Grey shading indicates a $\pm0.05$ threshold. Bottom: ideal ramp profile and reconstructed ramp profile for a linear ramp function offset from the gate centres by 50~nm. Centre: Gate potentials $U_j$ for the reconstructed upper ramp profile (blue solid line and filled circles) and lower ramp profile (red solid line and filled squares). Green (gray) shaded bands indicate position and extent of gates.\label{fig:ramp_reconstruction}}
\end{figure}

We quantify the fluctuations in the potential with the ratio
\beq
\Delta (W,M)&=& \frac{V_{\rm max}-V_{\rm min}}{V_{\rm max}+V_{\rm min}},
\eeq
which, if the oscillations were sinusoidal, would give the maximum absolute deviation relative to the mean. Fig.~\ref{fig:gates}c shows the regions in the $W$-$M$ plane for which $\Delta<5\%$ and $\Delta<2.5\%$. The bottom left of the figure, corresponding to more, narrower gates, reproduces the constant potential more accurately. In the main text, each leg of the nanowire device was taken to have a length of $2~\mu{\rm m}$. With gate dimensions as in Fig.~\ref{fig:gates}a, each leg can therefore support approximately $N$ gates, where
\beq
N&=&\frac{2000}{2(M+W)}.
\eeq
We superimpose contours of constant $N$ on Fig.~\ref{fig:gates}c to indicate how varying $W$ and $M$ alters the number of gates per leg. In particular, we can achieve a fluctuation ratio of $\Delta<5\%$ using $W=M=25~{\rm nm}$, or approximately 20 gates per leg.

However, we must also check that this gate arrangement can reproduce the varying part of the ideal potential profile. To do this, we take $W=M=25~{\rm nm}$ and try varying the individual gate potentials $U_j$ to reproduce a linear ramp function. Explicitly, we write the total potential as
\beq
V_{\rm tot}(x;\{U_j\})&=&\sum_{j}V_{\rm g}(x;x_j,U_j)
\eeq
and try to reproduce the linear ramp function
\beq
\mu(x)&=&r\left[\beta(x-x_c)\right]
\eeq
(see Eq.~\eqref{eq:linear_ramp}) by minimising the (scaled) mean square error (MSE),
\beq
{\rm MSE} &=&\int\left[V_{\rm tot}(x;\{U_j\})-\mu(x)\right]^2\dd x.
\eeq
We carry out the minimisation by taking a series of eleven gates and performing gradient descent to find a minimum of the MSE as a function of $\{U_j\}$. As our starting point, we take each $U_j$ to be equal to the value of the ideal ramp function at the corresponding $x_j$.

Fig.~\ref{fig:ramp_reconstruction} shows the reproduced ramp function for a ramp starting at the midpoint of a gate (i.e. with $x_c=0$) and a ramp starting at the midpoint between neighbouring pairs of gates (i.e. $x_c=50$). Also shown is the ideal linear ramp function $\mu(x)$ in grey, along with a shaded region corresponding to $\mu(x)\pm0.05$ (which is the fluctuation tolerance we expect in the region of constant potential). In the first case, the gate arrangement is able to reproduce the linear ramp function within the tolerance band. However, in the second case, the reproduced profile lies slightly outside the tolerance band at the top and bottom of the ramp. Despite this, in both cases, the realistic gate arrangement is able to reproduce a potential profile that varies between zero and one in a roughly linear fashion. By continuously tuning the gate potentials $U_j$ between these optimal values (shown in the central panel of Fig.~\ref{fig:ramp_reconstruction}), the ramp function will move smoothly across the device as required to perform a braid.

Overall, this rough calculation suggests that approximately 20 gates would be needed per leg of the nanowire device to reproduce our ideal ramp function within a reasonable tolerance. Equivalently, the gates should have a width of approximately 50~nm and be separated by a gap of approximately 50~nm. An interesting avenue for further work would be to study the effects of the unwanted oscillations on the braiding process, and to see whether the 5\% tolerance used here is sufficient or even if it can be relaxed. Of course, a complete discussion of this issue, beyond the scope of this work, would also require a detailed study of the screening effects that take place within the nanowire device, as well as the variation of the chemical potential across the nanowire cross section.

\end{document}